\documentclass[manuscript]{aastex61}

\usepackage{url}
\usepackage{tabularx}
\usepackage{amsmath}
\usepackage{rotating}
\usepackage{multirow}
\shorttitle{SXR emission of flares}
\shortauthors{Sadykov et al.}

\begin{document}
	
\title{Statistical Properties of Soft X-ray emission of solar flares}

\correspondingauthor{Viacheslav M Sadykov}
\email{vsadykov@njit.edu}

\author{Viacheslav M Sadykov}
\affiliation{Center for Computational Heliophysics, New Jersey Institute of Technology, Newark, NJ 07102, USA}
\affiliation{Department of Physics, New Jersey Institute of Technology, Newark, NJ 07102, USA}
\affiliation{Bay Area Environmental Research Institute, Moffett Field, CA 94035, USA}
\affiliation{NASA Ames Research Center, Moffett Field, CA 94035, USA}

\author{Alexander G Kosovichev}
\affiliation{Center for Computational Heliophysics, New Jersey Institute of Technology, Newark, NJ 07102, USA}
\affiliation{Department of Physics, New Jersey Institute of Technology, Newark, NJ 07102, USA}
\affiliation{NASA Ames Research Center, Moffett Field, CA 94035, USA}

\author{Irina N Kitiashvili}
\affiliation{Bay Area Environmental Research Institute, Moffett Field, CA 94035, USA}
\affiliation{NASA Ames Research Center, Moffett Field, CA 94035, USA}

\author{Alexander Frolov}
\affiliation{Center for Computational Heliophysics, New Jersey Institute of Technology, Newark, NJ 07102, USA}
\affiliation{Department of Computer Science, Cornell University, Ithaca, NY 14853, USA}

\begin{abstract}
	We present a statistical analysis of properties of Soft X-Ray (SXR) emission, plasma temperature (T), and emission measure (EM), derived from GOES observations of flares in 2002-2017. The temperature and emission measures are obtained using the TEBBS algorithm~\citep{Ryan12a}, which delivers reliable results together with uncertainties even for weak B-class flare events. More than 96\% of flares demonstrate a sequential appearance of T, SXR, and EM maxima, in agreement with the expected behavior of the chromospheric evaporation process. The relative number of such flares increases with increasing the SXR flux maximum. The SXR maximum is closer in time to the T maximum for B-class flares than for $\ge$C-class flares, while it is very close to the EM maximum for M- and X-class flares. We define flares as ``T-controlled'' if the time interval between the SXR and T maxima is at least two times shorter than the interval between the EM and SXR maxima, and as ``EM-controlled'' if the time interval between the EM and SXR maxima is at least two times shorter than the interval between the SXR and T maxima. For any considered flare class range, the T-controlled events compared to EM-controlled events have: a) higher EM but lower T; b) longer durations and shorter relative growth times; c) longer FWHM and characteristic decay times. Interpretation of these statistical results based on analysis of a single loop dynamics suggests that for flares of the same class range, the T-controlled events can be developed in longer loops than the EM-controlled events.
\end{abstract}


\section{Introduction}
\label{Section:introduction}

	Solar flares are the strongest transient energy release phenomenon on the Sun. Understanding the underlying physical mechanisms and their general properties is important for development of flare models and space weather forecasts. According to the CSHKP model \citep{Carmichael64a,Sturrock66a,Hirayama74a,Kopp76a,Priest02a,Shibata11a}, which is considered as a standard flare model, charged particles accelerated in a coronal current sheet penetrate deep into the chromosphere and deposit their energy. As a result, the strongly heated chromospheric plasma expands upward, filling coronal loops. This phenomenon, known as ``chromospheric evaporation'', can also be caused by other energy release and transport processes such as heat conduction \citep{Antiochos78a}, Joule heating \citep{Sharykin14a}, or high frequency Alfv\'en waves \citep{Fletcher08a,Reep16a,Kerr16a}.
	
	Although different mechanisms can be responsible for chromospheric evaporation, the general sequence of physical processes is qualitatively similar. After flare starts, the energy release mechanism heats the dense chromospheric plasma until it reaches maximum temperature, $T_{max}$. During and after this process, the hot chromospheric plasma expands into the coronal loops. Once the energy release weakens, the evaporated plasma flow also weakens, and at some point the plasma starts condensing back into the chromosphere. At this time moment, the plasma reaches its maximum emission measure, $EM=\int{}n^{2}dV$.
	
	The X-ray Sensor \citep[XRS,][]{Bornmann96a} onboard the Geostationary Operational Environmental Satellite (GOES) series currently provides one of the longest continuous observations of solar activity. GOES measures the Soft X-ray (SXR) flux in two channels, 1-8\,{\AA} and 0.5-4\,{\AA}. The maximum emission in the 1-8\,{\AA} channel is traditionally used to define the flare class, which often serves as a measure of the flare strength. Thus it is important to understand what physical characteristics, such as the plasma temperature and emission measure, are represented by the GOES classes. Assuming a single-temperature plasma approximation, \citet{Thomas85a}, \citet{Garcia94a}, and \citet{White05a} developed a procedure to compute the temperature (T) and emission measure (EM) of the flare plasma based on the GOES SXR measurements. Studies of \citet{Feldman96a,Garcia88a,Ryan12a} demonstrated correlations of the GOES SXR flux maximum with the maximum temperature and emission measure during solar flares. Because the instrument response per unit emission measure \citep[Figure~7 of][]{White05a} is a monotonically increasing function of temperature, increasing both T and EM can increase the SXR flux in both GOES channels. Since the maximum temperature, $T_{max}$, is expected to occur before the maximum of emission measure, $EM_{max}$, one can suggest that the maximum of the GOES SXR 1-8\,{\AA} flux is likely to occur between the T and EM maxima. The timeline of events is illustrated in Figure~\ref{figure:eventsequence}.
	
	Figure~\ref{figure:RADYNsequence} illustrate the SXR 1-8\,{\AA} flux, its derivative, temperature, and emission measure calculated for the standard flare model using the RADYN radiative hydrodynamic code~\citep{Allred15a}. The model results are in qualitative agreement with the timeline in Fig.~\ref{figure:eventsequence}. The RADYN code solves \text{time-dependent} hydrodynamic, radiative transfer, and non-equilibrium atomic level population equations on an adaptive 1D vertical grid. Specifically, Figure~\ref{figure:RADYNsequence} shows the results for the RADYN model ``radyn\_out.val3c\_d3\_1.0e12\_t20s\_15kev\_fp'' from the F-CHROMA solar flare model database (\url{http://www.fchroma.org/}). In this model, the energy flux is deposited by a beam of accelerated electrons (with a 15\,keV low-energy cutoff and a power-law spectral index of 3) propagating downwards and delivering a total of $10^{12}$erg\,cm$^{-2}$ into the atmosphere in 20 seconds. The SXR emission is calculated for each grid point of the RADYN model assuming that the loop cross-section is $S=10^{18}$\,cm$^{2}$, and summed up over the grid points for each time snapshot. Using these SXR light curves and assuming a single-temperature plasma approximation, we calculate T and EM.
	
	The motivation of this work is to understand the relationships between the plasma parameters (maximum values of T and EM and the corresponding times) and the properties of the SXR emission (GOES class, emission duration, characteristic times, etc). In particular, we define events as ``T-controlled flares'' if the SXR maximum -- T maximum time interval is at least two times shorter than the EM maximum -- SXR maximum interval, and as ``EM-controlled flares'' if the EM maximum -- SXR maximum time interval is at least two times shorter than the SXR maximum -- T maximum interval, and try to answer the following questions:
	\begin{enumerate}
		\item How often do flares obey the scenario of the chromospheric evaporation process illustrated in Figure~\ref{figure:eventsequence}?
		\item Which of the plasma properties, T or EM, mainly influence the SXR maximum value and timing for the different GOES flare classes?
		\item What is the physical difference between the T-controlled and EM-controlled flares of the same GOES class range?
	\end{enumerate}
	
	The implemented algorithm to calculate T and EM and the event selection process are described in Section~\ref{Section:dataselection}. Relationships among the flare characteristics are presented in Section~\ref{Section:results}, followed by a discussion in Section~\ref{Section:discussion}. A short summary of the results and conclusions is presented in Section~\ref{Section:conclusion}.

\section{Data selection and processing}
\label{Section:dataselection}

	To estimate behavior of T and EM during solar flares, we have applied the Temperature and Emission measure-Based Background Subtraction algorithm \citep[TEBBS,][]{Bornmann90a,Ryan12a}, which allows the user to obtain T and EM values for flares detected by the GOES satellite. In this algorithm, the background level of the GOES X-ray emission is taken into account in order to obtain T and EM during the whole flare duration, including the rising phase. Our Python realization of the TEBBS algorithm for coronal element abundances is available at \url{https://github.com/vsadykov/TEBBS}. We note that the GOES data allow us to determine the flare temperature and emission measure only in a single-temperature approximation, because the data are obtained only in two SXR energy channels.
	
	In this work, we analyze the GOES data obtained from January, 2002 to December, 2017. The full list of events for this time period is obtained from the Interactive Multi-Instrument Database of Solar Flares \citep[IMIDSF, ][\url{https://helioportal.nas.nasa.gov}]{Sadykov17a}. In the TEBBS algorithm, we add the Savitzky-Golay smoothing procedure with third-order polynomials and a 30-second running window function \citep{Savitzky64a}. This allows us to perform analysis for relatively weak B-class flares and also compute smooth derivatives of the light curves.
	
	For each event, we determine the maximum values and times of the temperature, emission measure, and background-subtracted SXR 1-8\,{\AA} flux and its derivative. We determine the flare end time as the moment when the background-subtracted flux drops by a factor of two from its maximum value. This definition is the same as in the GOES flare catalog but instead of the preflare SXR flux level we subtract the derived background. The resulting difference in the flare end times does not change our conclusions. Then, we calculate various parameters related to the flare temporal behavior: the duration, growth time (defined as the maximum time minus start time), relative growth time (defined as the growth time divided by the flare duration), and the time intervals among the maxima of T, EM, SXR, and SXR derivative. Following \citet{Reep17a} we calculate the FWHM of the GOES 1-8\,{\AA} light curve (defined as the difference between two time moments when the SXR reaches the half maximum value before and after its maximum) and its characteristic decay time, $\tau{}_{decay}$, at the flare end defined as:
	
	\begin{gather}
	\label{eq:timedecay}
	\tau{}_{decay}=-\dfrac{F_{SXR}(t)}{dF_{SXR}(t)/dt}|_{t=t_{end}}
	\end{gather}
	
	In addition, when possible, we define the flare ribbon areas from the catalog by \citet[][for the events of the SDO epoch only]{Kazachenko17a}. We exclude from our analysis the events, for which:
	\begin{itemize}
		\item The background-subtracted SXR maximum flux is lower than the flux of B1.0 class flare ($10^{-7}$\,W\,m$^{-2}$) because of the low S/N ratio;
		\item The maximum temperature shows the presence of the super-hot plasma \citep[$>$30\,MK,][]{Caspi14a}, indicating that the single-temperature model is not valid \citep{Sharykin15a};
		\item The relative uncertainties of the maximum temperature or emission measure are greater than 100\%;
		\item The maximum of the emission measure occurs when the flux in the GOES 0.5-4\,{\AA} channel is smaller than 1\% of the flare maximum flux in this channel (i.e. almost at the level of the background);
		\item The gaps in the GOES data are longer than 5\% of the flare duration;
		\item The TEBBS algorithm does not return any reliable background combination (for all combinations (1) the number of rising phase time bins is low, (2) the flare maximum temperature does not exceed 3\,MK \citep{Ryan12a}, or (3) the preflare T and EM values are higher than their maximum values during the flare);
	\end{itemize}
	
	We found that a total of 14955 out of 22728 flares satisfy these criteria. The flares, which do not satisfy the above criteria and are excluded from the consideration, are mainly weak B-class or C-class events. The final statistical sample includes 5915 B-class, 7774 C-class, 1159 M-class, and 107 X-class flare events. Hereafter, we consider the X-class flares in one group together with the M-class flares.

\section{Results}
\label{Section:results}

	\subsection{Time sequence of events during solar flares}
	
		We compare the order of appearance of the maxima of T, EM, and SXR flux, as well as their derivatives, with the flare scenario illustrated in Figure~\ref{figure:eventsequence}. First, we found that 94.5\% of B-class flares (5587 out of 5915), 97.3\% of the C-class flares (7568 out of 7774), and 98.6\% of M- and X-class flares (1248 out of 1266) follow the sequential appearance of the T, SXR, and EM maxima, i.e. the assumed chromospheric evaporation scenario. On average, 96.3\% of all flares follow the sequence, and the fraction of such flares increases with the GOES class.
		
		Second, we found that for 82.5\% of all flare events (82.6\% of B-class flares, 83.4\% of C-class flares, and 76.7\% of M- and X-class flares) the 1-8\,{\AA} SXR derivative maximum mainly occurs prior the T maximum. Interestingly, the fraction of such events does not increase with the GOES class, and even becomes lower for M- and X-class flares.

	\subsection{Physical parameters controlling the SXR emission for flares of different classes}
	
		In order to understand which of the two parameters, T or EM, determines the timing of the maximum of the SXR emission, we consider the time intervals between the SXR and T maxima and between the EM and SXR maxima. Two-dimensional diagrams of the SXR-T and EM-SXR intervals for different flare classes are presented in Figure~\ref{figure:tplots}. White horizontal and vertical dashed lines restrict the zones where events do not obey the sequential T, SXR, and EM maxima appearances. Inclined white dashed lines represent places in the histogram where one of the intervals is two times longer than another.
		
		The relationships are very different for the different flare classes. Among the B-class flares (Figure~\ref{figure:tplots}a), 34.0\% are T-controlled and 26.4\% are EM-controlled. The situation is completely opposite for the M- and X-class flares (Figure~\ref{figure:tplots}c): the number of EM-controlled events is 82.8\% and just 1.7\% are T-controlled. The C-class flares (Figure~\ref{figure:tplots}b) fall between these two cases: 43.6\% events are EM-controlled and 18.0\% are T-controlled. We can see that the SXR maximum occurs very close to the EM maximum mostly for the M- and X-class flares and closer to the T maximum for the B-class flares. In this respect, the weak and strong flares behave differently.
		
		Figures~\ref{figure:BCselected}a~and~\ref{figure:BCselected}b show the dependence of the T and EM maximum values from the SXR maximum flux. It is essentially the same as previously presented by \citet{Ryan12a}. As one can see, both the temperature and emission measure maxima are correlated with the SXR maximum, and the correlations for the EM are more prominent. For M- and X-class events (with the logarithm of the maximum flux of -5.0 or greater) the logarithm of the EM maximum is proportional to the logarithm of the SXR flux maximum. 
		
		Figure~\ref{figure:BCselected}c presents the relationship between the flare SXR class as defined in the GOES flare catalog and the SXR maximum flux calculated after subtraction of the background. The background subtraction is especially important for the B-class and low C-class events, because, as is evident from the figure, the same SXR flux corresponds to a wide spread of the flare classes in the GOES catalog.
		
		Figure~\ref{figure:BCselected}d illustrates the relationship between the SXR maximum flux, and the time interval between the temperature and the SXR derivative maxima. As one can see, the temperature reaches the maximum in most cases after the derivative maximum. The median value of the time interval slightly varies with the SXR flux maximum, and for strong $\geq$M-class flares becomes smaller than for B-class and C-class flares (for the flares with SXR flux maxima of $<10^{-5}$\,W\,m$^{-2}$), although it does not exceed one minute for any flare class.

	\subsection{Difference between T-controlled and EM-controlled events.}
	
		Figure~\ref{figure:tplots} reveals a transition in behavior of the time intervals with an increase in the flare's SXR maximum flux. One should expect that in some range of the SXR maximum fluxes the number of the T- and EM-controlled events is almost the same. We have found that such situation happens when the logarithm of the SXR flux maximum (in W\,m$^{-2}$) is between -6.20 and -5.80 (i.e. for the background-subtracted SXR classes of B6.3~- C1.6). Figure~\ref{figure:tplots}d illustrates the time interval relationship for this range. Among the selected flares, 1365 are EM-controlled, and 1176 are T-controlled. The B6.3~- C1.6 class range is indicated in Figure~\ref{figure:BCselected} by white vertical dashed lines.
		
		Previously, we have concluded that the relationship between the SXR-T and EM-SXR time intervals depends on the SXR flux maximum. By selecting relatively narrow class ranges we can study the influence of other physical parameters to the partition among the intervals. Table~\ref{table1_TEBBSparameters} summarizes the median values and corresponding median absolute deviations of physical parameters for the EM-controlled and T-controlled flares in several such class ranges. The median absolute deviation is a dispersion measure of a univariate data set, and is equal to a median of absolute deviations of the data points from the data set median. One can see the following trends from this table. Except for the $\ge$M1.0 class range (due to insufficient statistics), the T-controlled events are cooler on average than the EM-controlled events and have larger emission measures. Also, these events are typically longer in duration, but grow faster (have shorter relative growth times) and have greater FWHM and characteristic decay times.
		
		Figure~\ref{figure:1dhistos}a-f presents histograms of various physical parameters for the T-controlled (red) and EM-controlled (gray) flare events of the B6.3~- C1.6 class range for the same characteristics as in Table~\ref{table1_TEBBSparameters}. As one can see, although the distributions significantly overlap, they prominently differ from each other. The difference is especially visible for the relative growth times, where two peaks of the histogram are clearly separated. The histograms constructed for other class ranges (B1.0~- B2.5, B2.5~- B6.3, C1.6~- C4.0, and C4.0~- M1.0) are found to have similar behaviors.

\section{Discussion}
\label{Section:discussion}

	In this Section we summarize our answers to the questions posted in Section~\ref{Section:introduction} and discuss possible explanations.

\begin{figure}[h]
	\centering
	\includegraphics[width=1.0\linewidth]{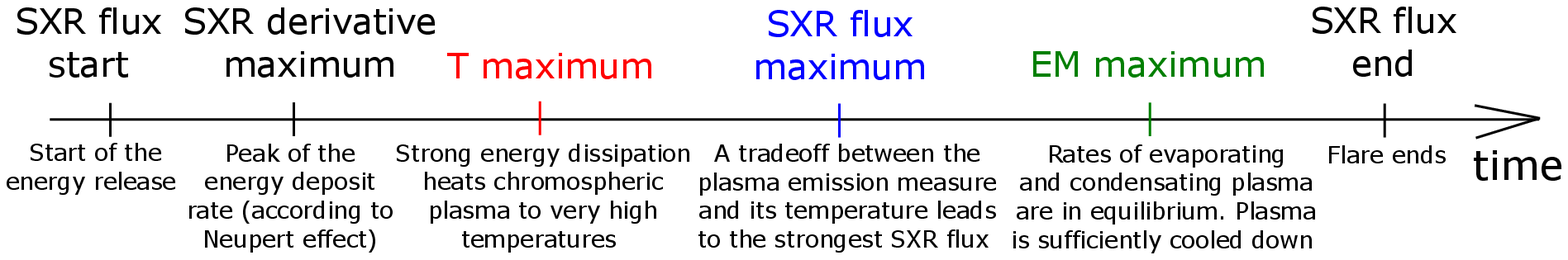}
	\caption{Time sequence of events during the chromospheric evaporation process in solar flares.}
	\label{figure:eventsequence}
\end{figure}

\begin{figure}[h]
	\centering
	\includegraphics[width=1.0\linewidth]{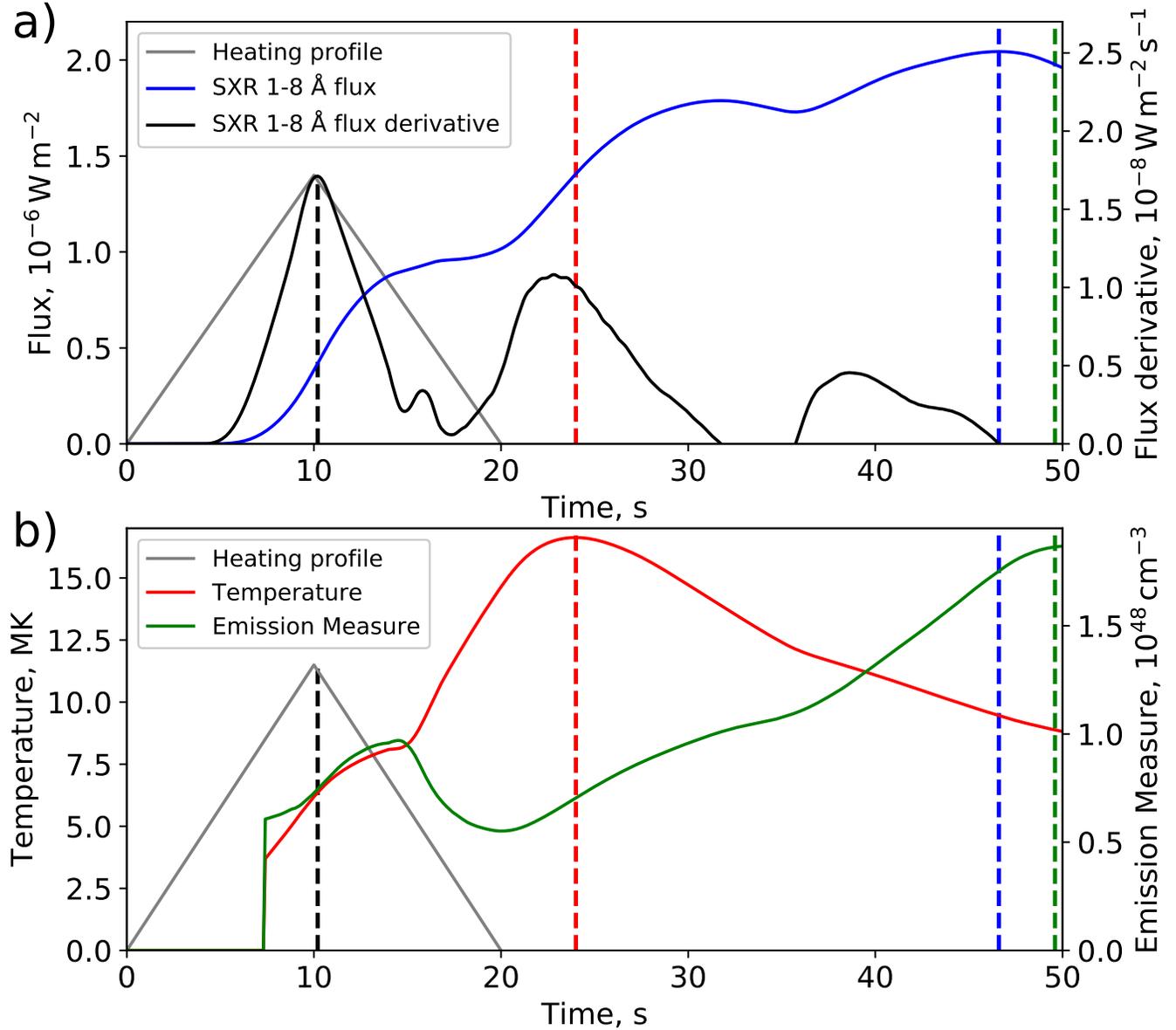}
	\caption{(a) SXR 1-8\,{\AA} flux and its derivative calculated for RADYN model ``radyn\_out.val3c\_d3\_1.0e12\_t20s\_15kev\_fp'' from the F-CHROMA solar flare model database (\url{http://www.fchroma.org/}); (b) temperature and emission measure calculated from the modeled SXR 0.5-4\,{\AA} and 1-8\,{\AA} fluxes. Loop cross-section of $S=10^{18}$\,cm$^{2}$ is assumed for these calculations. Gray triangle represents the deposited energy flux profile. Dashed vertical lines mark the maxima of presented characteristics.}
	\label{figure:RADYNsequence}
\end{figure}

\begin{figure}[h]
	\centering
	\includegraphics[width=0.49\linewidth]{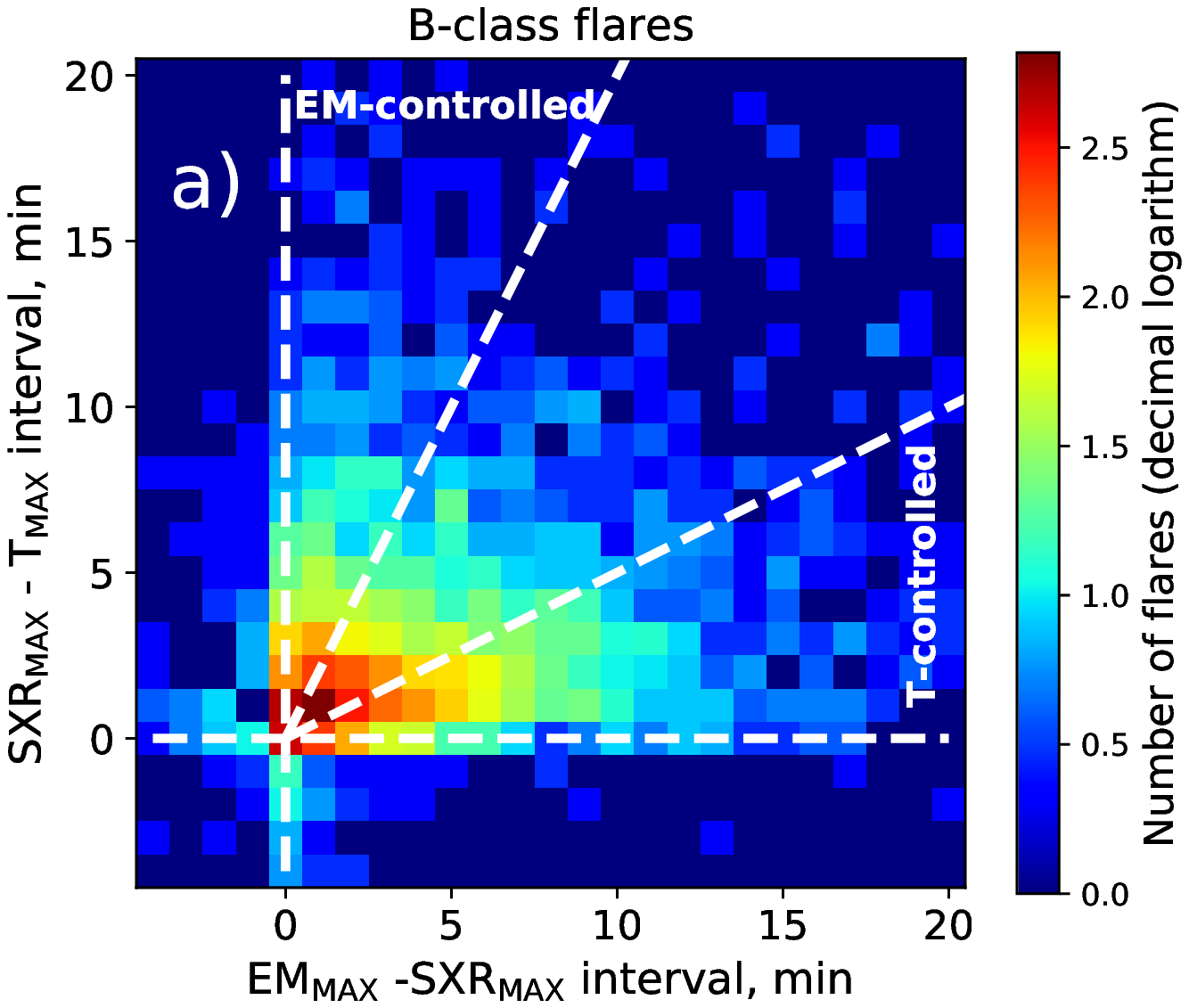}
	\includegraphics[width=0.49\linewidth]{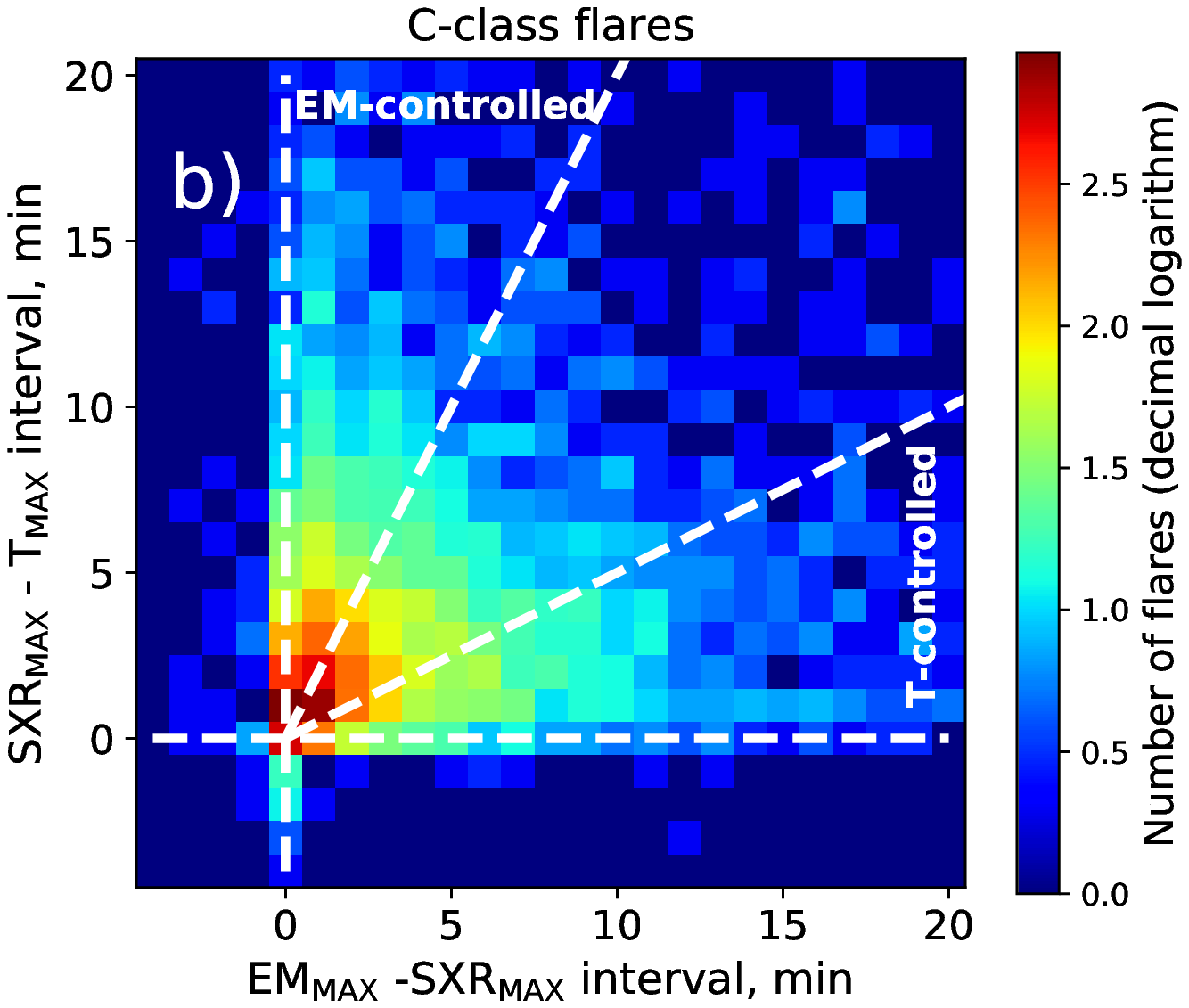} \\
	\includegraphics[width=0.49\linewidth]{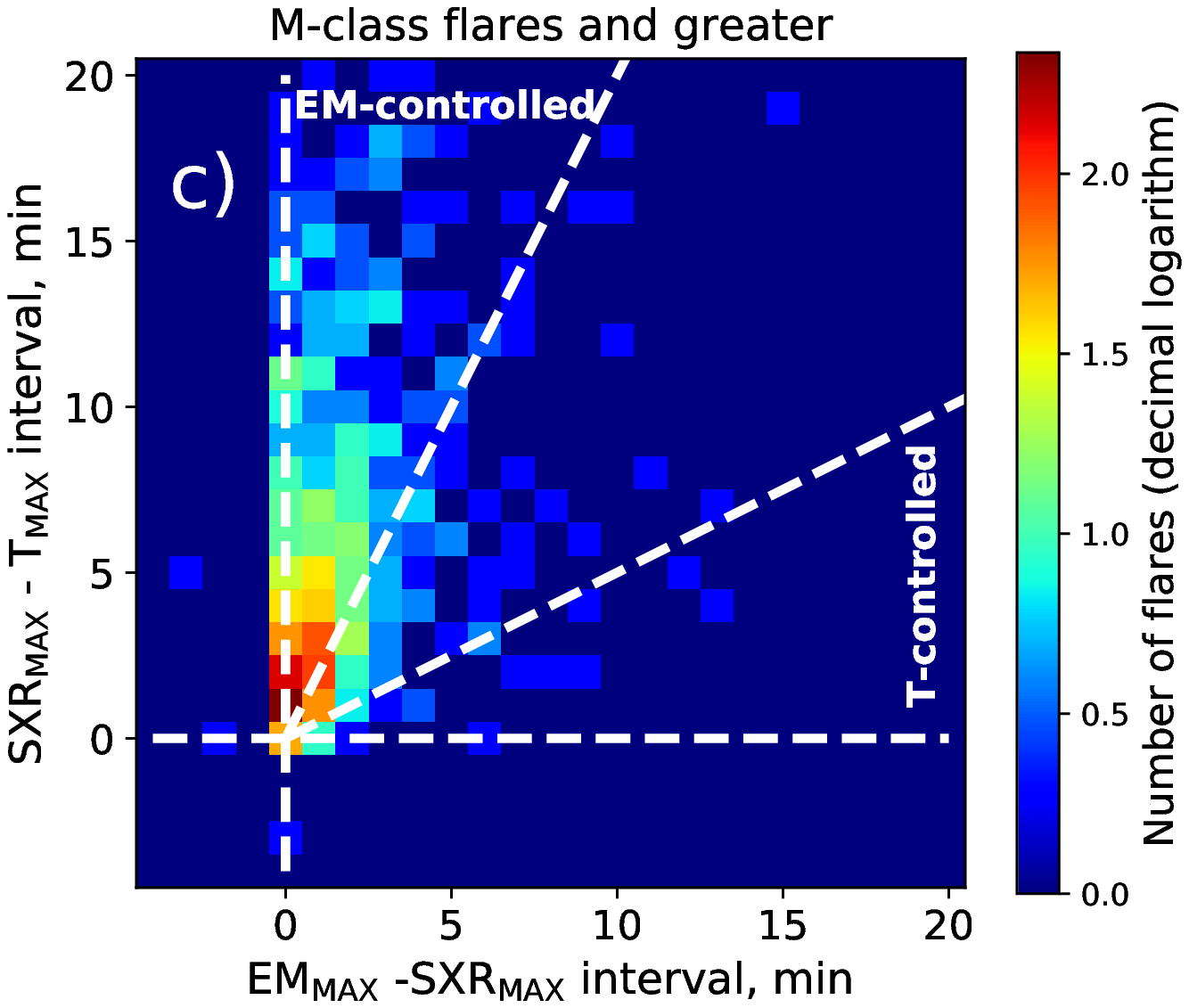}
	\includegraphics[width=0.49\linewidth]{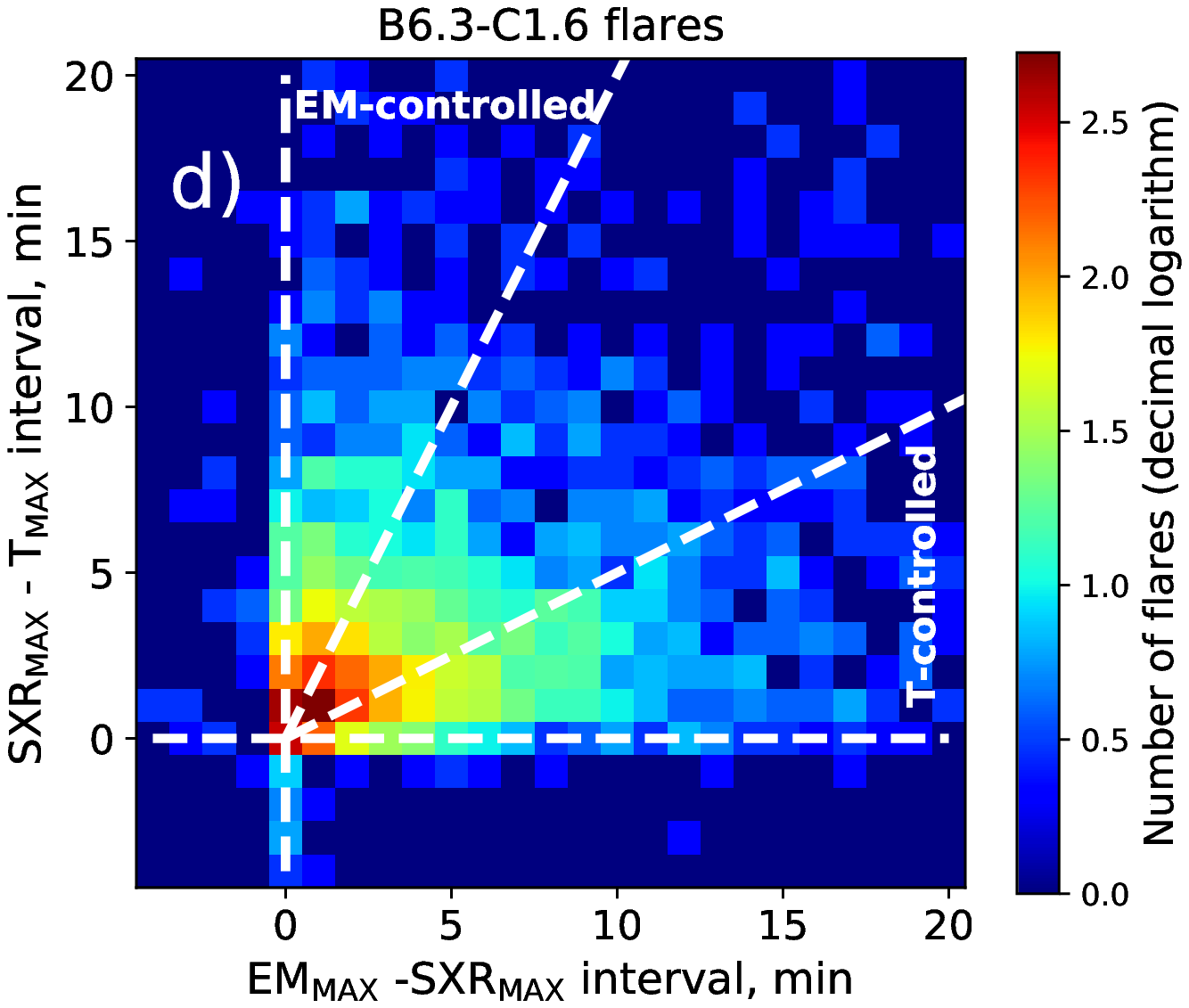} \\
	\caption{Two-dimensional relationships of the time intervals between the EM and SXR maxima (x-axis) and SXR and T maxima (y-axis) for a) B-class flares, b) C-class flares, and c) M- and X-class flares. White dashed lines show zones for the T-controlled events and the EM-controlled events. Panel (d) illustrates the same relationships for the flares of B6.3~- C1.6 classes.}
	\label{figure:tplots}
\end{figure}

\begin{figure}[h]
	\centering
	\includegraphics[width=0.49\linewidth]{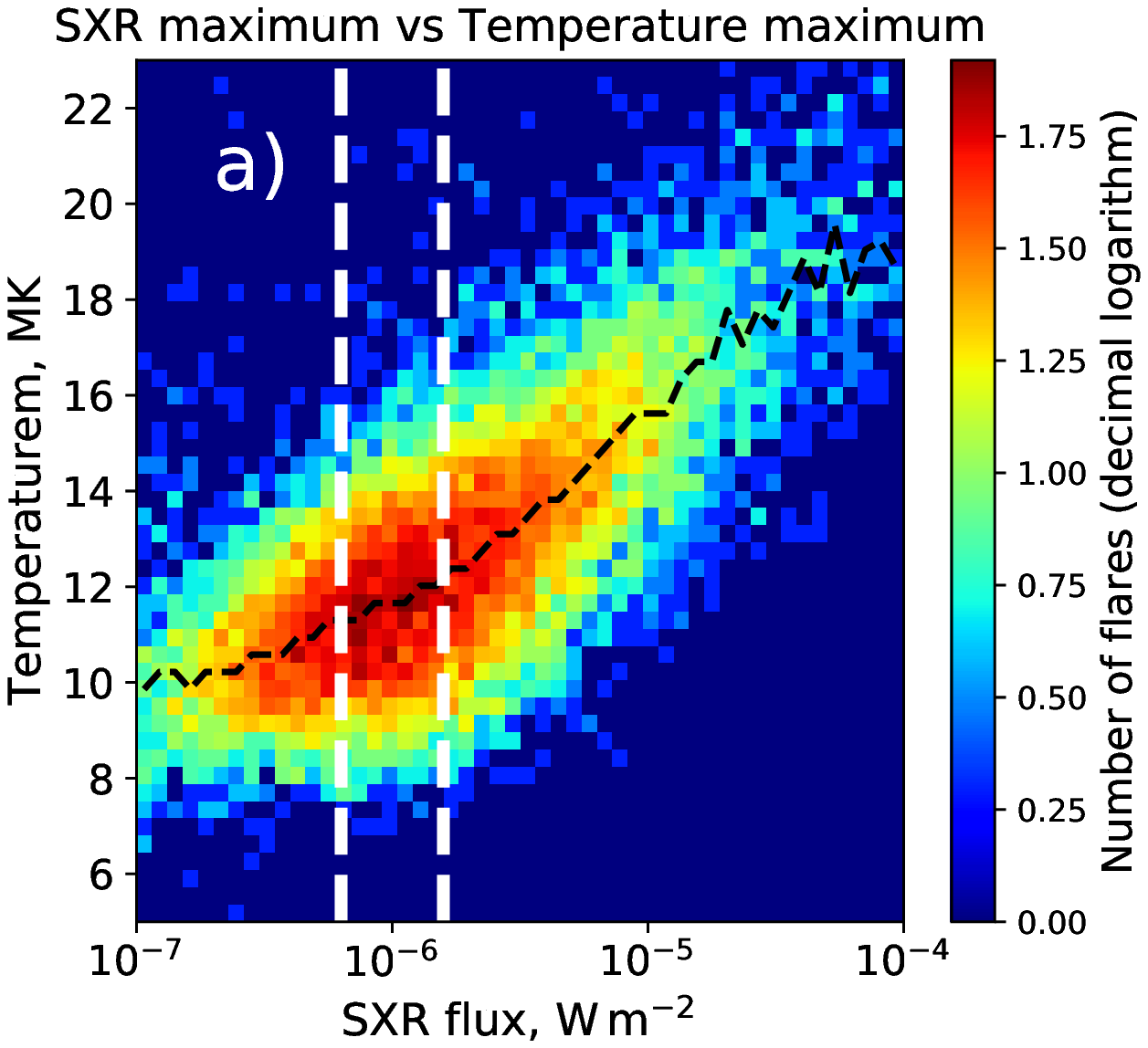}
	\includegraphics[width=0.49\linewidth]{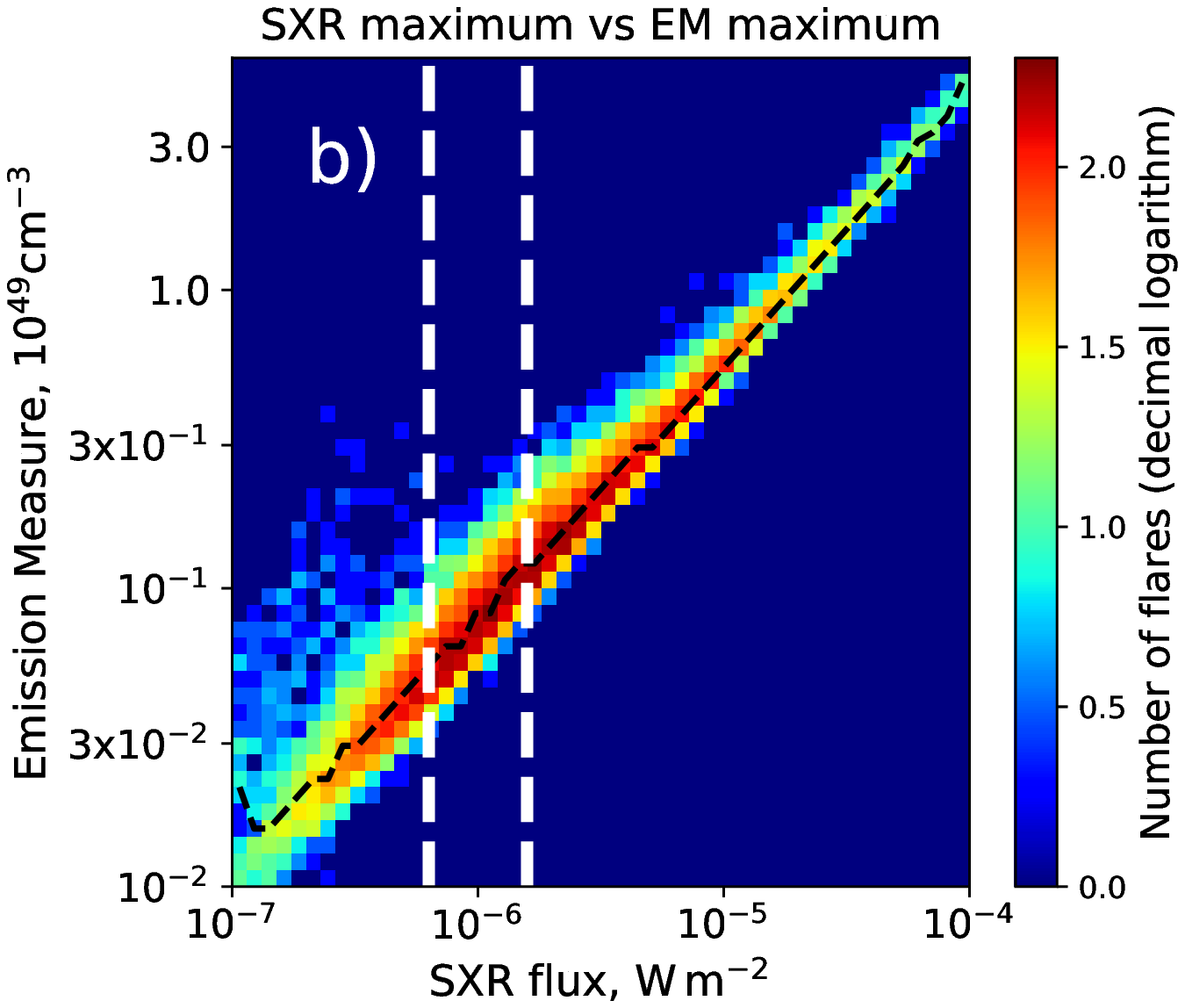} \\
	\includegraphics[width=0.49\linewidth]{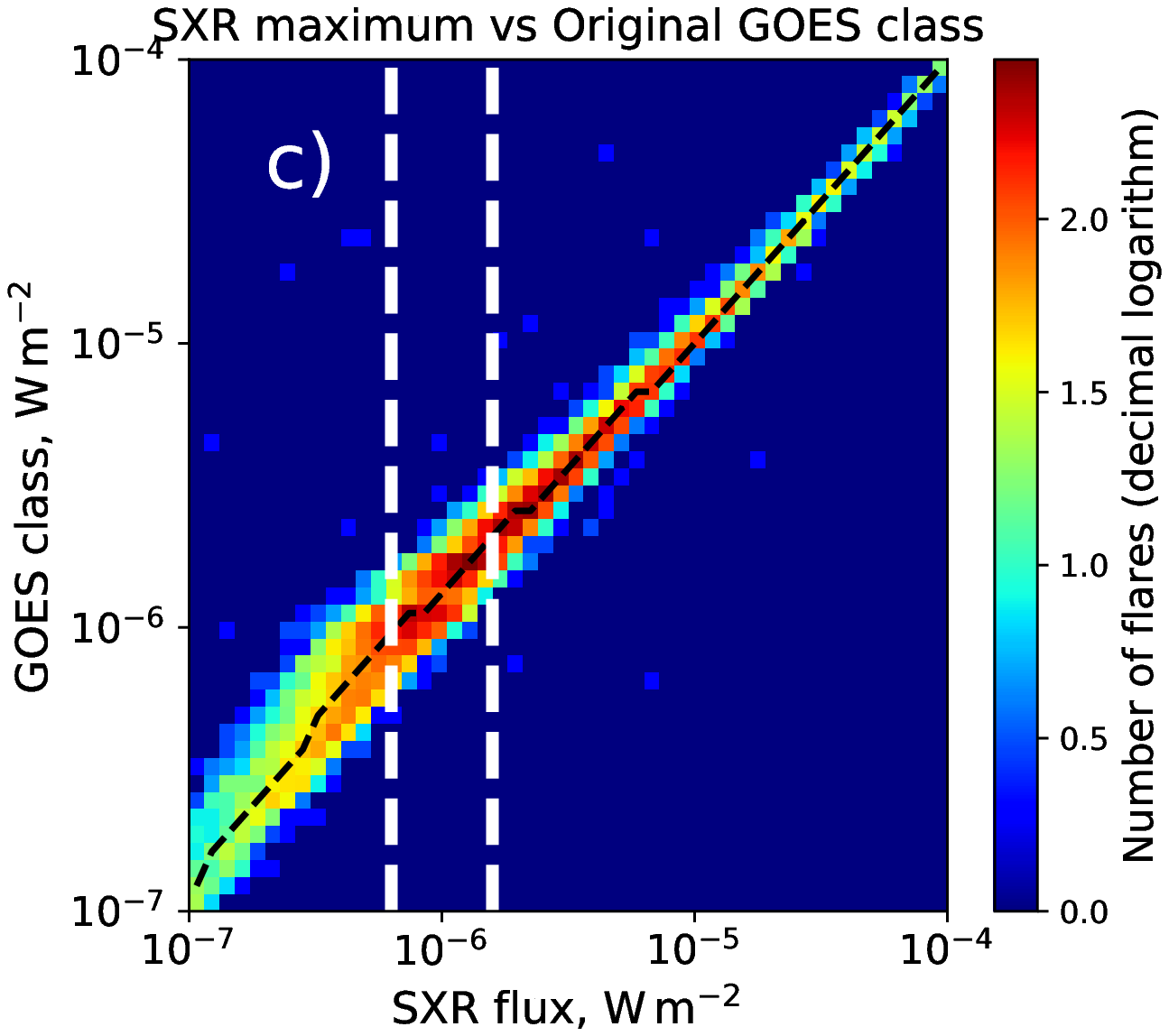}
	\includegraphics[width=0.49\linewidth]{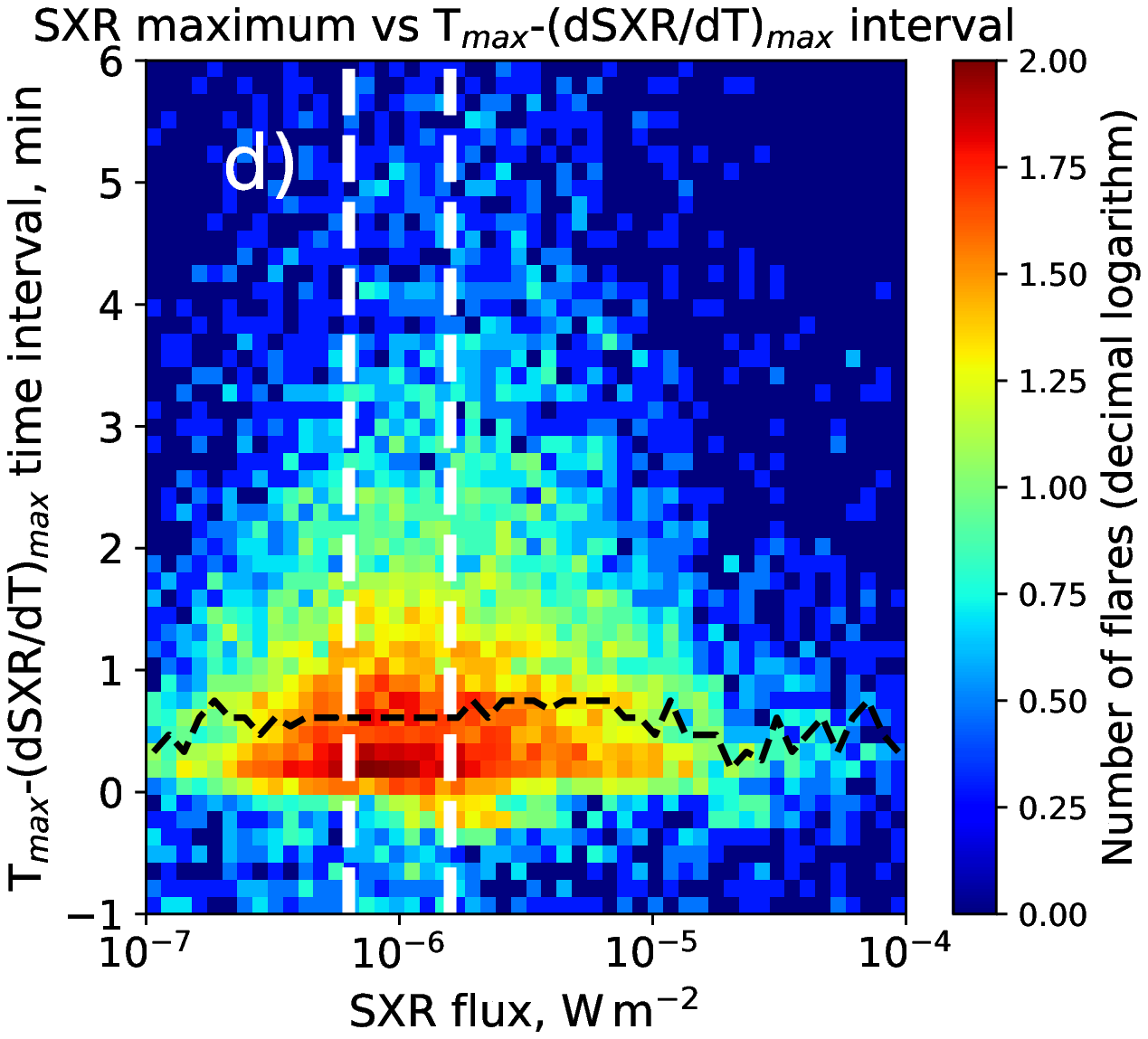} \\
	\caption{Two-dimensional relationships of (a) temperatures, (b) emission measures, (c) GOES classes, and (d) time intervals between the T and SXR derivative maxima of flare events, and their SXR maximum fluxes. Black dashed lines show median values for each SXR maximum flux. White vertical lines mark the B6.3~- C1.6 flare class range selected for the detailed study.}
	\label{figure:BCselected}
\end{figure}

	\subsection{How often do the flares obey the scenario of the chromospheric evaporation process illustrated in Figure~\ref{figure:eventsequence}?}
	
		We found that the temperature (T), soft X-ray flux (SXR), and emission measure (EM) maxima appear sequentially during solar flares for most of analyzed events (96.3\% on average). The fraction of such events increases with increasing SXR flux maximum. The observed sequence of the temperature, soft X-ray emission, and emission measure maxima fits into the standard picture of chromospheric evaporation, supported by the radiative hydrodynamic simulations (Fig.~\ref{figure:eventsequence}). However, what is the reason for the remaining 3.7\% (552 out of 14955) of events not following the sequence? Most of these events are weak B-class flares. Possibly, the T and EM calculations for weak events are not sufficiently accurate because of the relatively strong background level during these events.
		
		Another interesting fact is that, for most flares, the SXR derivative maximum occurs even before the temperature maximum. According to the Neupert effect \citep{Neupert68a}, the derivative of the SXR flux are correlated with the time of the energy deposit of high-energy electrons. Therefore, for most of events the strongest energy deposit happens before the plasma reaches the highest temperature, because some time is required for the deposited energy to heat the plasma. This is in agreement with the considered RADYN simulation (Figures~\ref{figure:eventsequence}b~and~\ref{figure:eventsequence}c). Note that the fraction of such events decreases with the flare class. We do not try to explain this in the present study.

	\subsection{Which of the plasma properties, T or EM, mainly influence the SXR maximum value and timing for the different GOES flare classes?}
	
		Figure~\ref{figure:tplots} illustrates that the temperature, emission measure, and SXR 1-8\,{\AA} flux light curves behave differently for different class flares. For the weak B-class flare events we see that the SXR maximum predominantly occurs closer to the T maximum than to the EM maximum. For the stronger M- and X-class flares, the SXR maximum occurs very close to the emission measure maximum. Also, Figure~\ref{figure:BCselected}b illustrates that the logarithm of the SXR maximum is proportional to the logarithm of the EM maximum for strong flares. Thus, one can conclude that the GOES class of strong M- and X-class flares most often represents the emission measure of the evaporated plasma and not the plasma temperature.
		
		In contrast, for the weak B-class flares we see that the SXR maximum is close to the T maximum. However, Figure~\ref{figure:tplots}a (illustrating the relations between the T and SXR maxima) does not show any direct relations between these two parameters for weak B-class flares. This means that for the weak events the corrected-for-background GOES class depends on both the temperature and emission measure.

	\subsection{What is the physical difference between the T-controlled and EM-controlled flares of the same GOES class range?}
	
		Our results indicate that for relatively narrow class ranges (see Table~\ref{table1_TEBBSparameters}) the T-controlled events are colder, have larger maximum EMs, are longer in duration, have shorter relative growth times, and have longer FWHMs as well as longer decay times. One of the possibilities is that the high-temperature plasma in the T-controlled events evolves in longer magnetic loops than in the EM-controlled events.
		
\begin{figure}[h]
	\centering
	\includegraphics[width=0.32\linewidth]{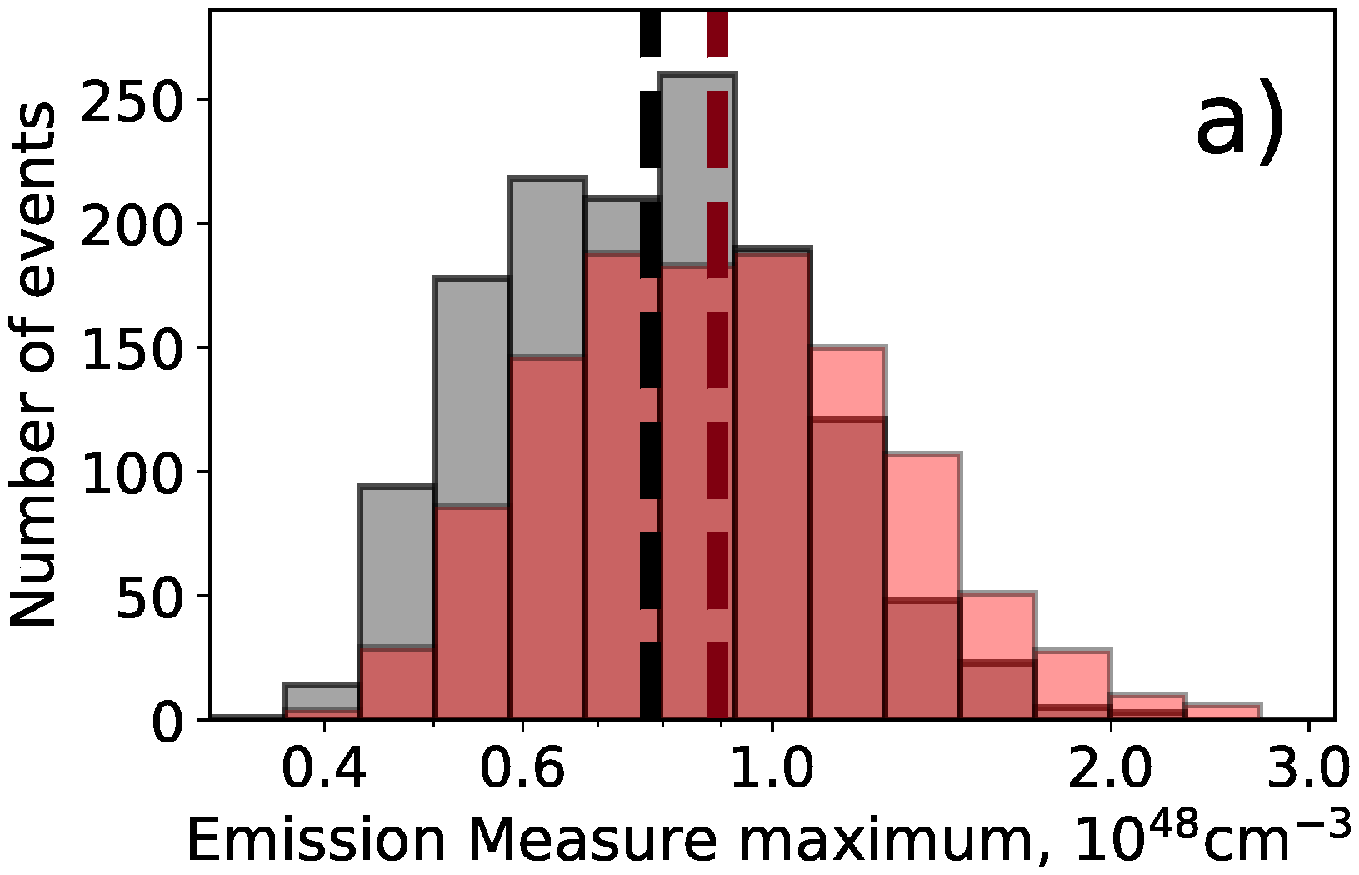}
	\includegraphics[width=0.32\linewidth]{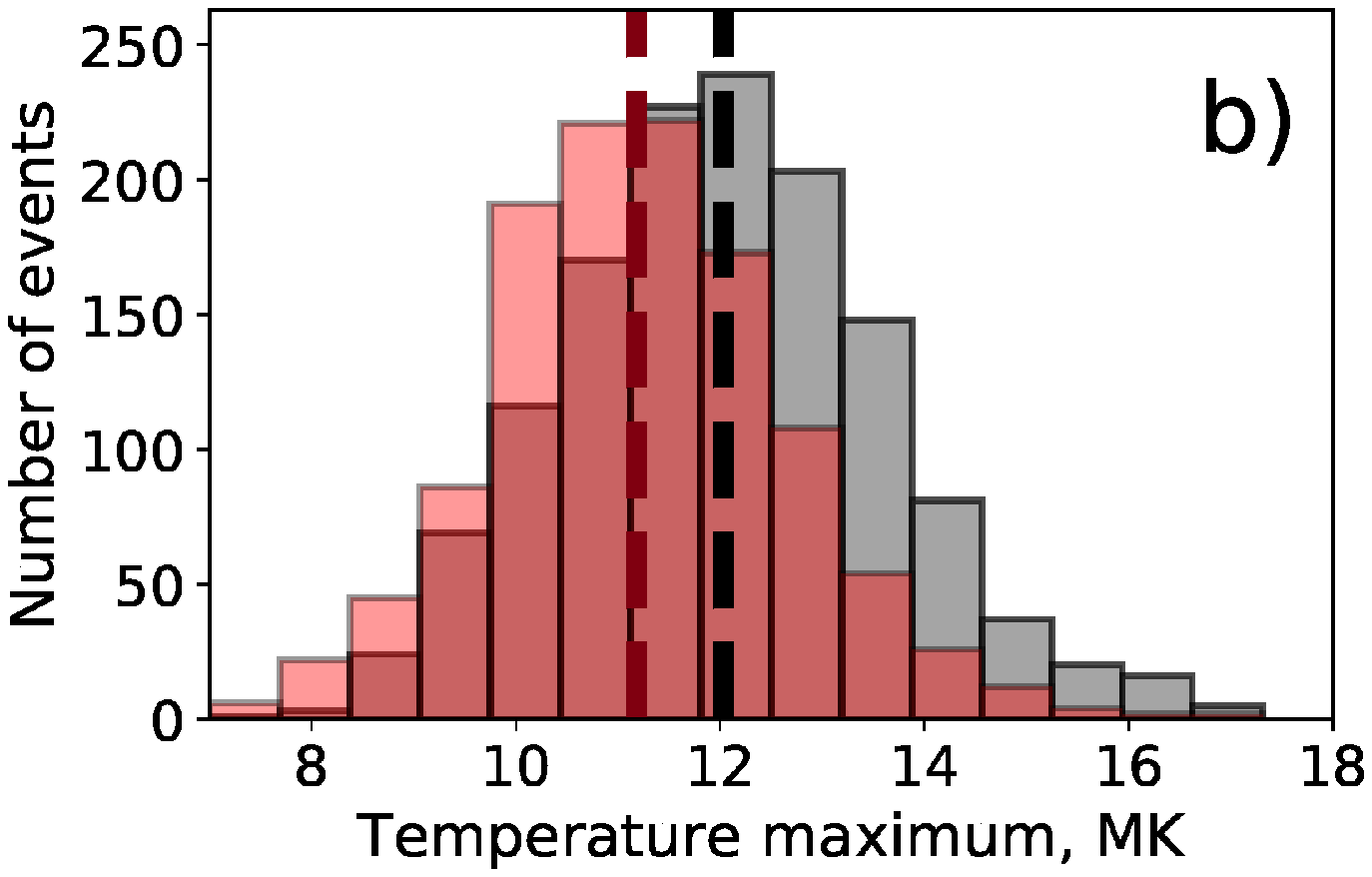}
	\includegraphics[width=0.32\linewidth]{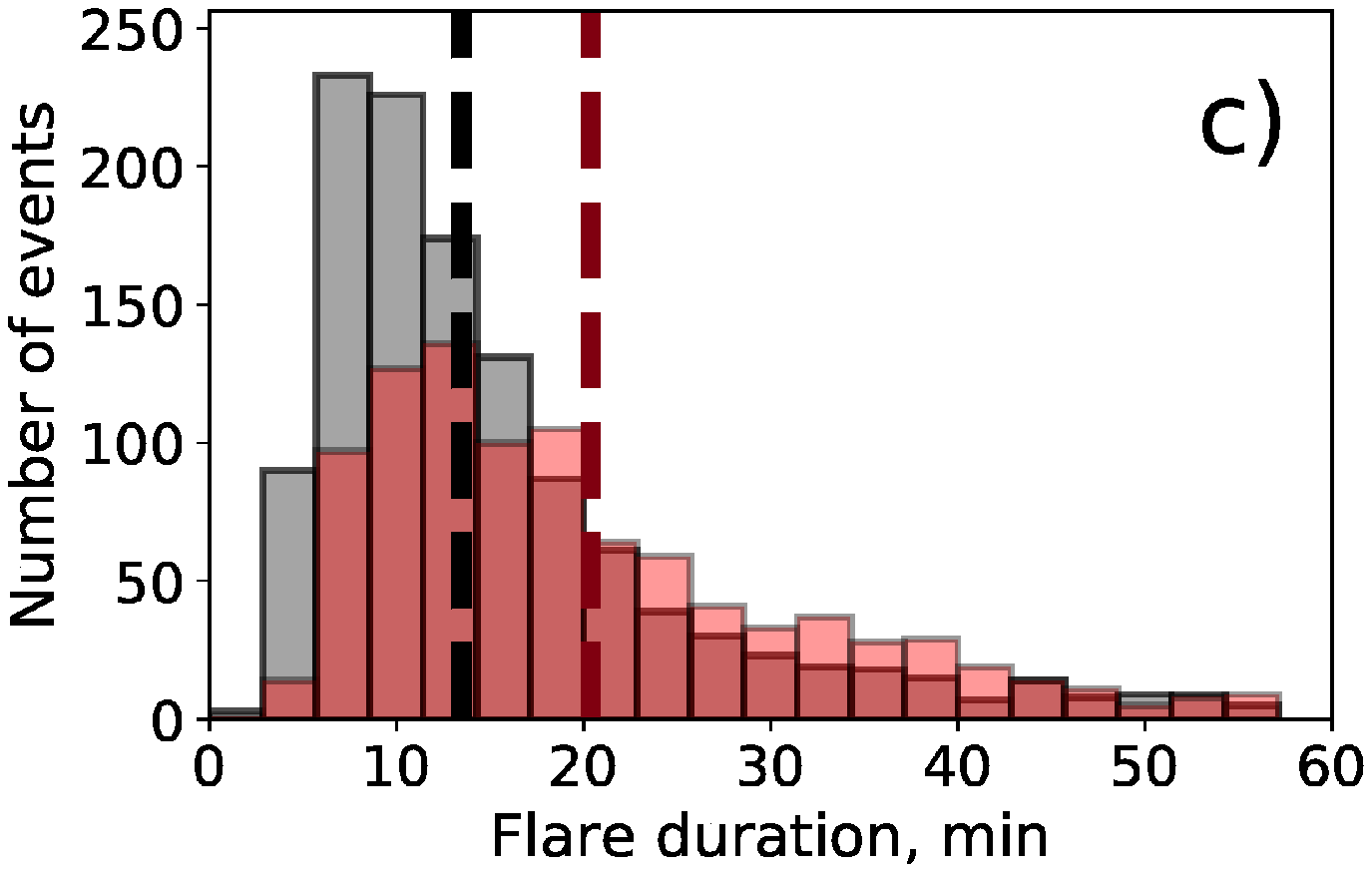} \\
	\includegraphics[width=0.32\linewidth]{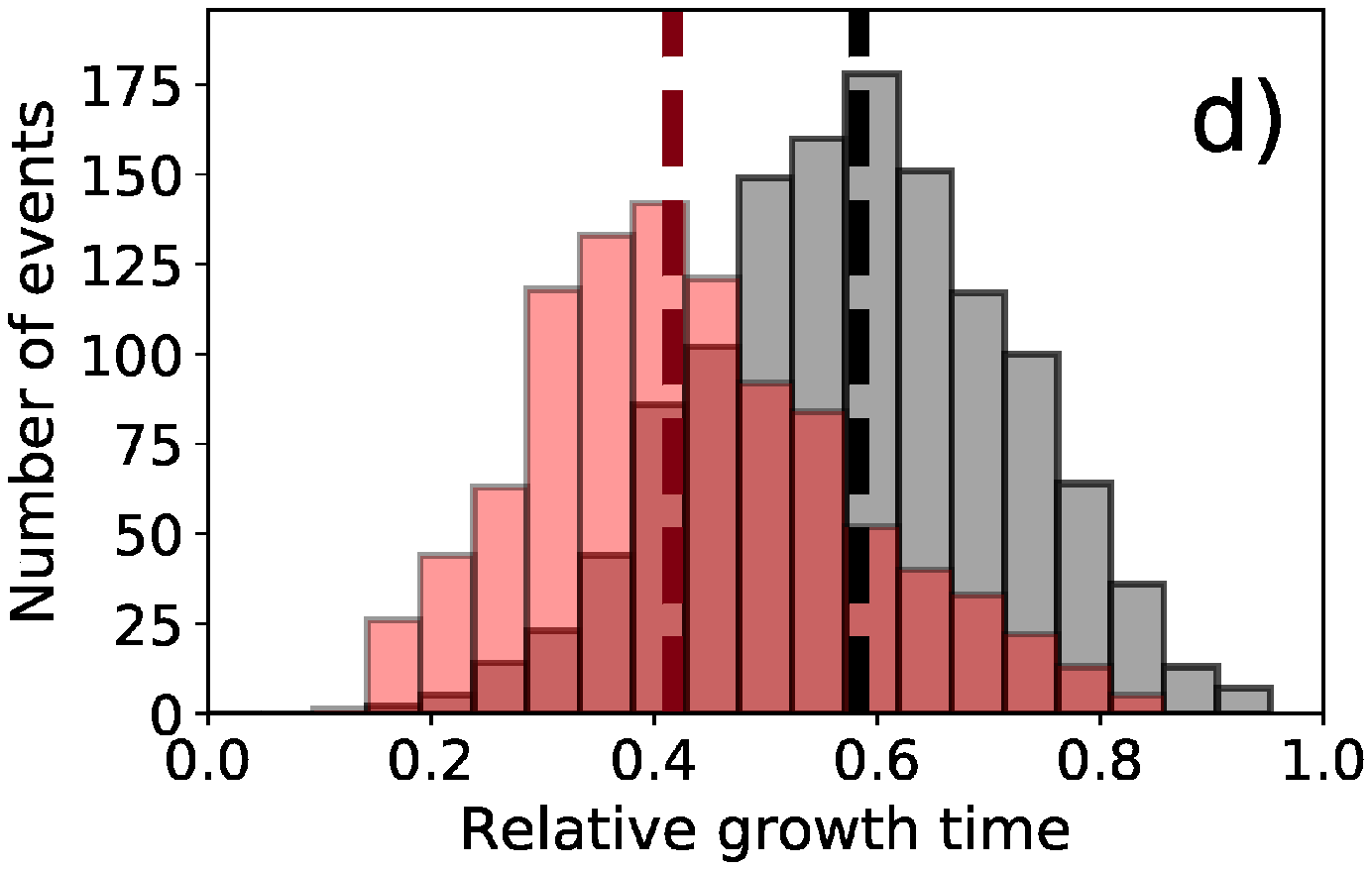}
	\includegraphics[width=0.32\linewidth]{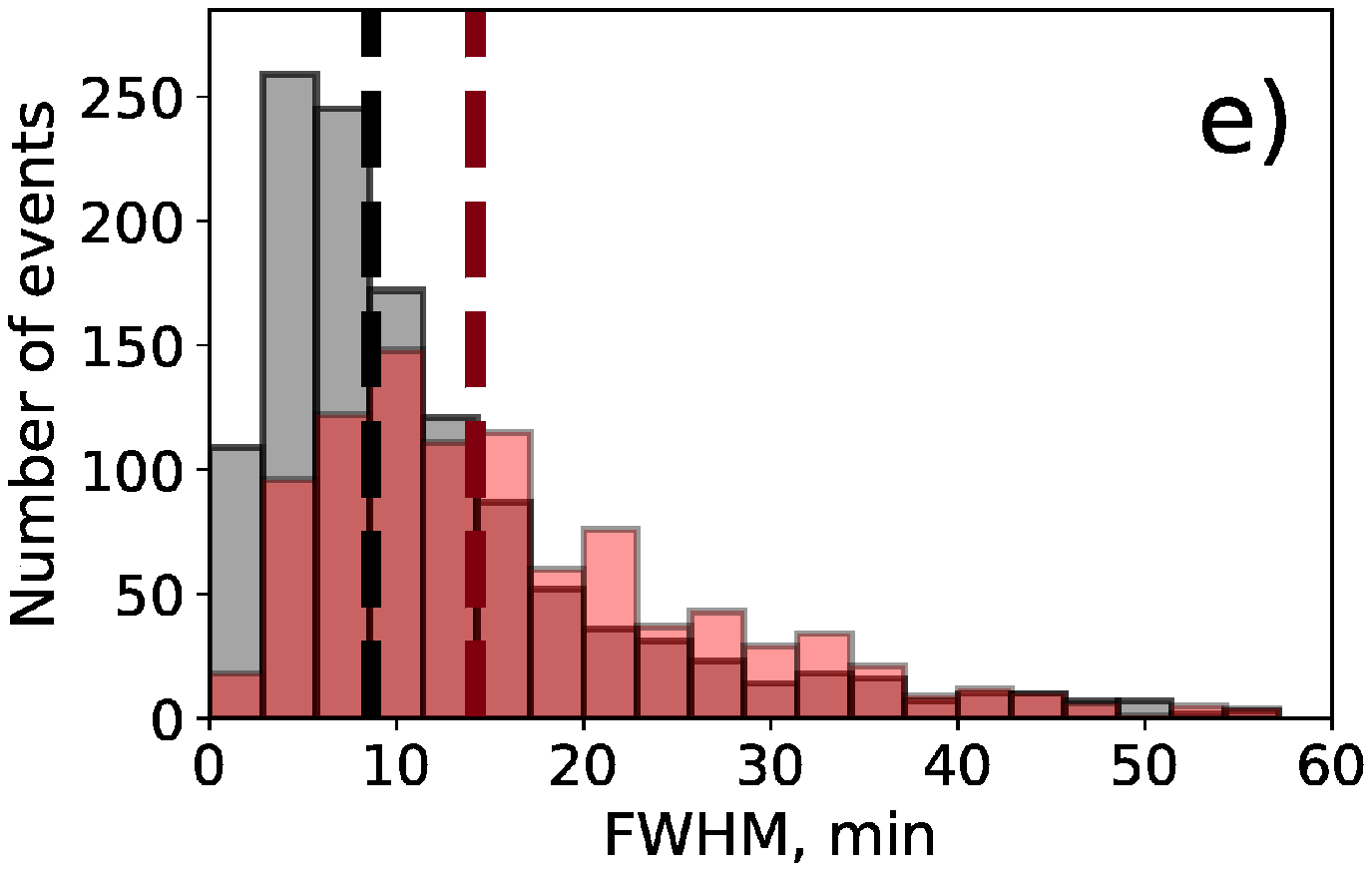}
	\includegraphics[width=0.32\linewidth]{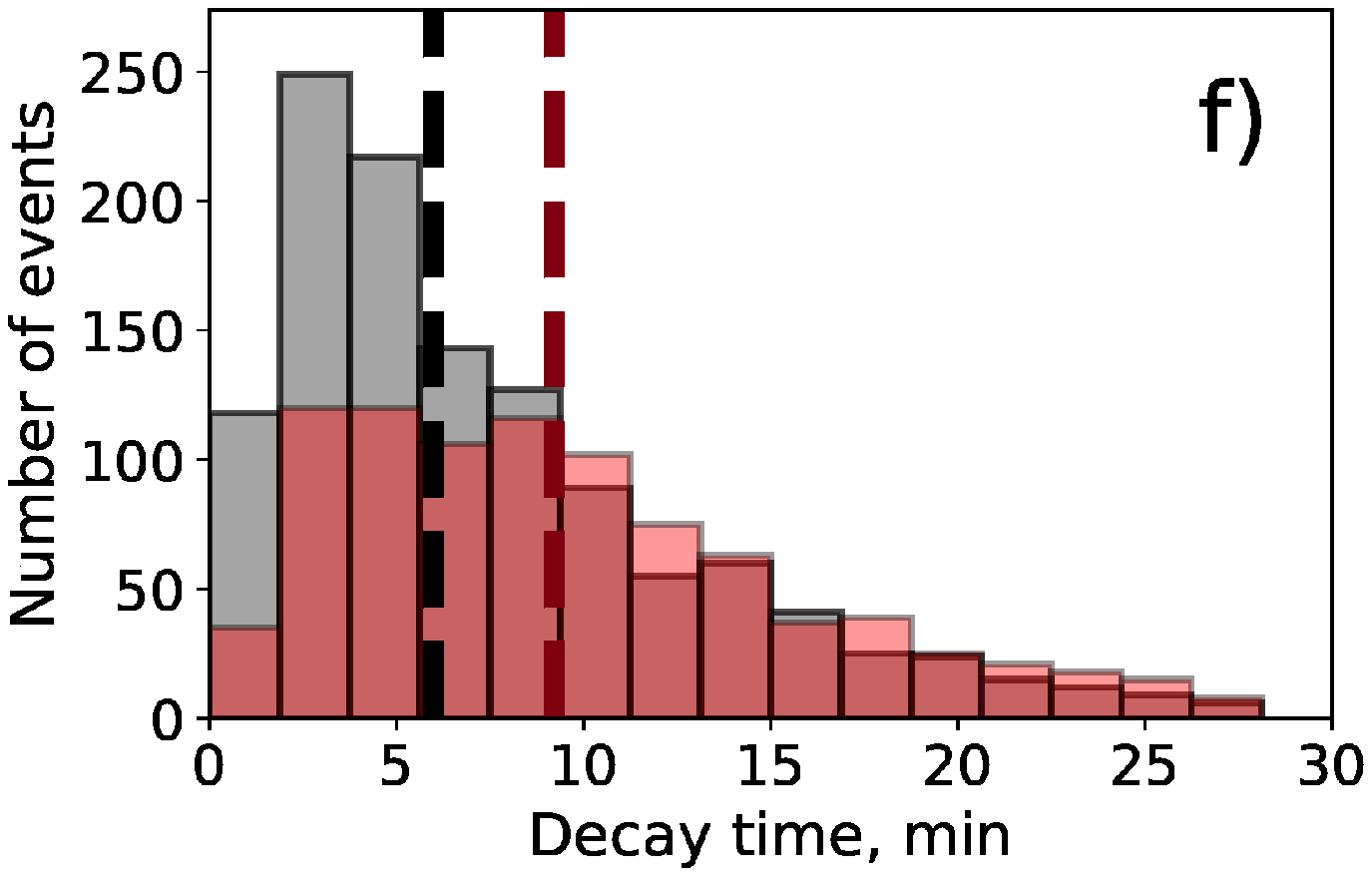} \\
	\includegraphics[width=0.32\linewidth]{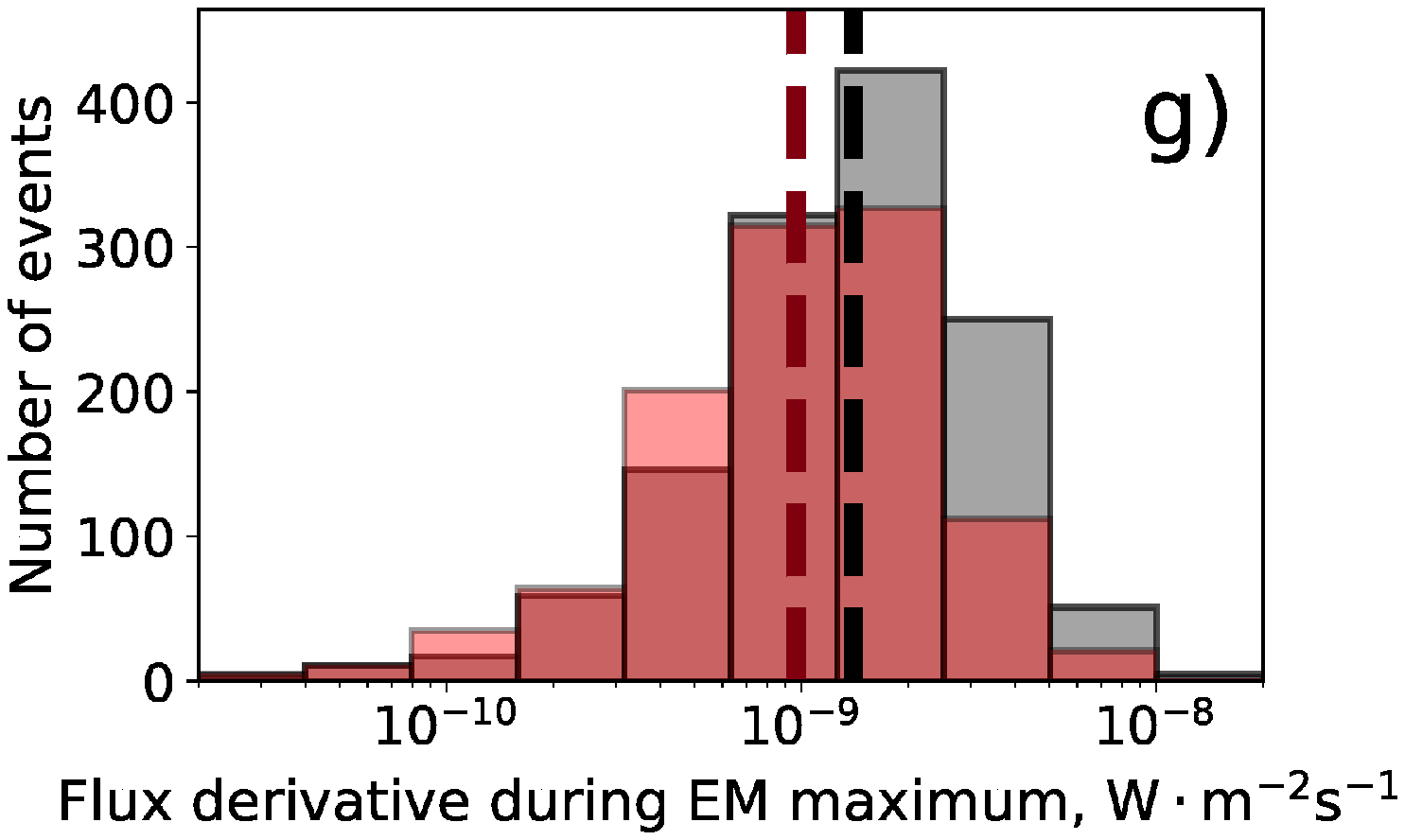}
	\includegraphics[width=0.32\linewidth]{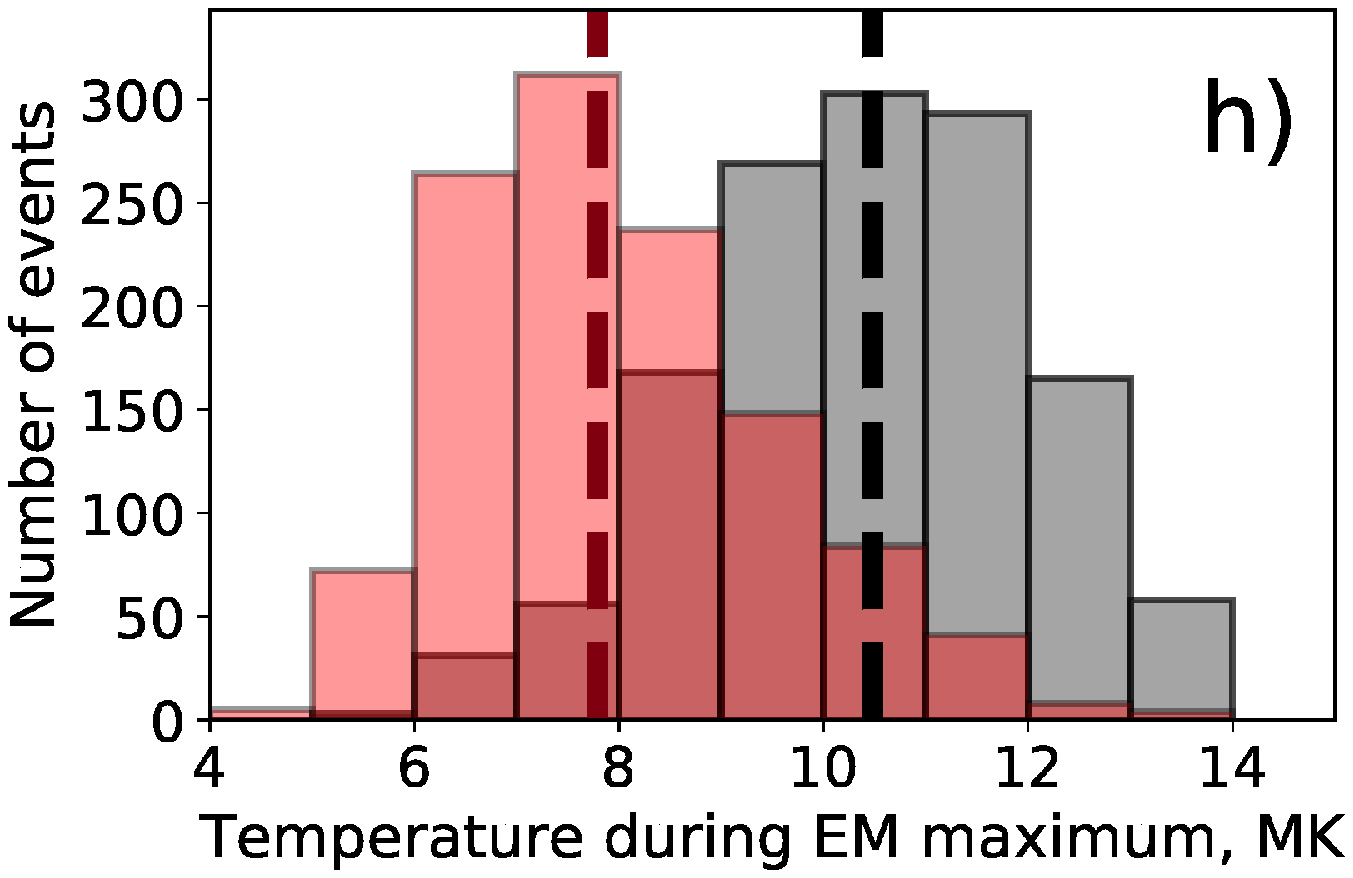}
	\includegraphics[width=0.32\linewidth]{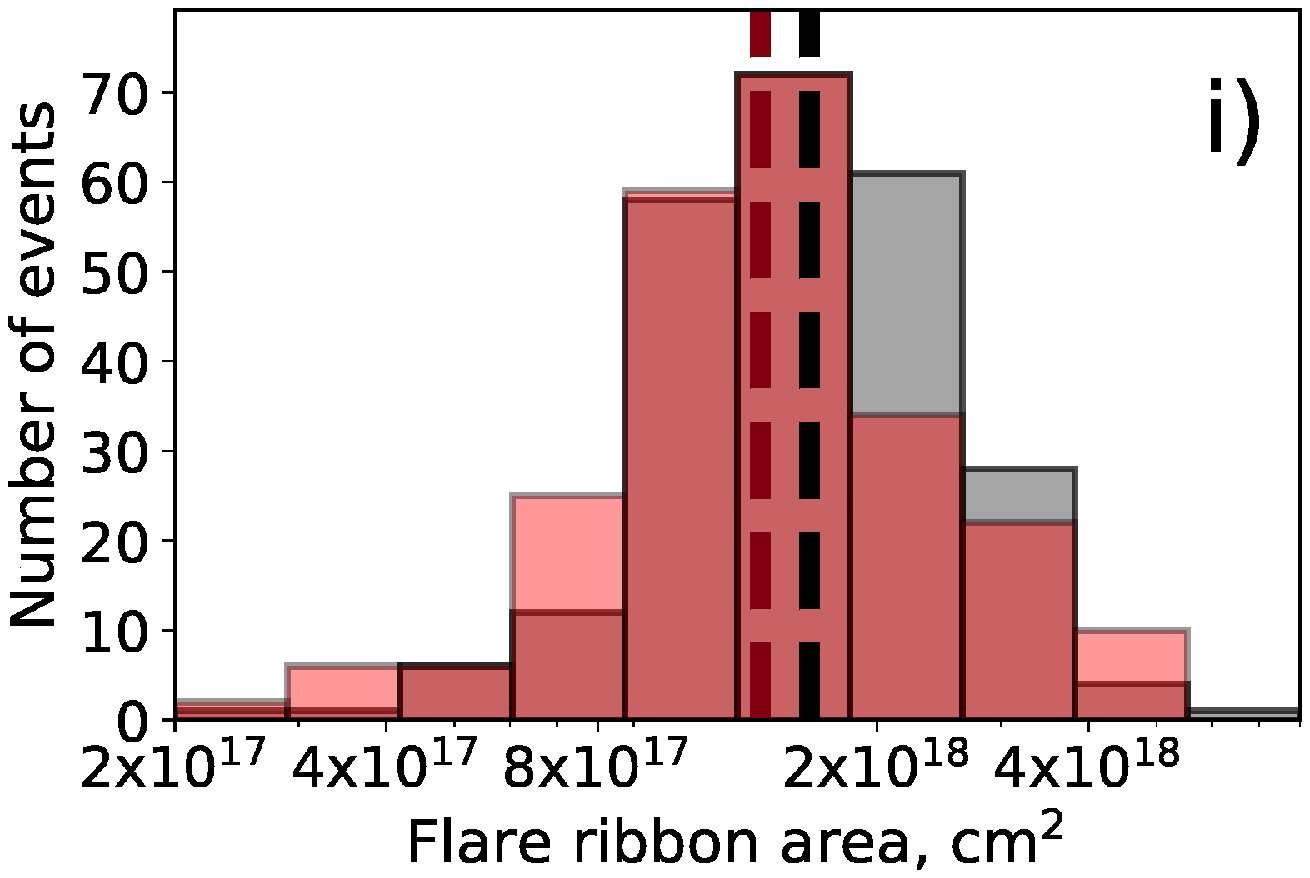} \\
	\caption{Histograms for the T-controlled (red) and the EM-controlled (gray) events of (a) EM maxima, (b) T maxima, (c) duration of the event, (d) event growth time (from SXR start to maximum time) normalized to the duration of the event, (e) FWHM, (f) characteristic decay time, (g) the SXR derivative during the EM maximum, (h) temperature during the EM maximum, and (i) ribbon areas of the events, for the flares of B6.3~- C1.6 GOES classes. Red dashed vertical line represents the median value for T-controlled events, black~--- for EM-controlled events. The corresponding median absolute deviations are presented in Table~\ref{table1_TEBBSparameters}.}
	\label{figure:1dhistos}
\end{figure}

		Several previous studies point to this conclusion. For example, \citet{Bowen13a} performed a statistical study of 17 flares of the $\approx$C8 class simultaneously observed by GOES, SDO/AIA, and SDO/EVE. The authors found that the flares with longer durations are usually cooler and evolve in larger volumes. As was mentioned before, the T-controlled events are on average longer and have lower temperatures than the EM-controlled flares and thus should evolve in larger volumes to maintain the observed EM level. Figure~\ref{figure:1dhistos}i illustrates the flare ribbon areas for events of the B6.3~- C1.6 class range from the flare ribbon catalog \citep{Kazachenko17a}. One can see that the ribbon areas on average are almost the same for the T-controlled and EM-controlled events. Thus, the larger volumes with the same ribbon areas will correspond to the longer loops. Several works based on the classical relations for coronal loop parameters \citep{Rosner78a} predicted longer decay times for the events evolving in longer loops~\citep{Serio91,Aschwanden08b}. As seen in Table~\ref{table1_TEBBSparameters}, the characteristic decay times are longer for the T-controlled events than for the EM-controlled events.
		
		\citet{Reep17a} provided direct observational and modeling relations between the GOES SXR light curve parameters (FWHM and characteristic decay times) and the distances between the flare ribbons. One of the conclusions of this work is that the events with longer FWHM and decay times should have larger separation of flare ribbons, i.e., evolve in longer loops. Although the observational relations in \citet{Reep17a} are derived for the flares of M\,5.0 GOES class or higher, the modeled SXR emission linearly scales with the assumed cross-section area, which makes it possible to extend the modeling relations to any GOES classes. The T-controlled events from our study have longer FWHMs and decay times and thus should evolve in longer loops according to the conclusion of \citet{Reep17a}.
		
		For further interpretation, we analyze the dynamics during the flare decay phase using the analytical equations for the spatially-averaged plasma parameters in magnetic loops implemented in Enthalpy-Based Thermal Evolution of Loops model \citep[EBTEL,][]{Klimchuk08a,Cargill12a,Cargill12b}. Following Eq.~2 in \citet[][]{Cargill12a}:
		
		\begin{gather}
		\label{eq:EBTEL-gen}
			\dfrac{L}{2(\gamma{}-1)}\dfrac{dp}{dt} = \dfrac{\gamma}{\gamma{}-1}p_{0}v_{0} - F_{0} + \dfrac{L}{2}Q - \dfrac{L}{2}n^{2}\Lambda{}(T)
		\end{gather}
	
		Here $p$, $n$, $T$ are the pressure, number density, and temperature averaged along the loop; $\gamma$ is the adiabatic constant, $L$ is the length of the loop, $F_{0} = \kappa{}_{0}T^{5/2}\dfrac{\partial{}T}{\partial{s}}|_{s=0}$ is the heat conduction flux at the loop footpoint, $Q$ is the heating term, and $\Lambda{}(T)$ is the radiative loss function depending on the temperature. We consider the loop energetics during the EM maximum time for the four flare class ranges (B2.5~- B6.3, B6.3~- C1.6, C1.6~- C4.0, and C4.0~- M1.0), for which we have statistics for flares with defined ribbon areas (last column in Table~\ref{table1_TEBBSparameters}). The B6.3~- C1.6 class properties presented in Figure~\ref{figure:1dhistos}g-i show that the T-controlled events have slightly lower SXR 1-8\,{\AA} derivative and lower temperatures during the EM maximum, and almost the same flare ribbon areas compared to the EM-controlled events. The median values of these parameters for the considered flare class ranges are summarized in Table~\ref{table2_looplengths}.
		
		We assume that during the emission measure maximum, the inflow and outflow of plasma into the flare volume are balanced, $p_{0}v_{0} = 0$, there is no heating along the loop at that time, i.e. $\dfrac{L}{2}Q=0$, and there is no change of the flare volume, $\dfrac{dEM}{dt}=0\Rightarrow{}\dfrac{dn}{dt}=0$. Following \citet{Rosner78a,Aschwanden08b,Aschwanden09a}, we approximate the radiative loss function in the 2-40\,MK temperature range as $\Lambda{}(T)\sim{}10^{-17.73}T^{-2/3}=\Lambda{}T^{-2/3}$. The equation of state is taken to be $p=nk_{B}T$ . The emission measure is $EM=n^{2}LS$, where $S$ is the cross-sectional area of the loop. Then, from Eq.~\ref{eq:EBTEL-gen} we obtain:
		
		\begin{gather}
		\label{eq:EBTEL-red1}
		\dfrac{dT}{dt} = \dfrac{2(\gamma{}-1)}{k_{B}L}\sqrt{LS/EM}(-F_{0} - EM\dfrac{\Lambda{}}{2ST^{2/3}})
		\end{gather}
		
		Following \citet{White05a}, the GOES flux in 1-8\,{\AA} channel can be represented as $F_{1-8{\AA}} = C\times{}EM\times{}\phi{}(T)$, where $\phi{}(T)$ is the function corresponding to the coronal element abundances shown in lower panel of Figure 7 of \citet{White05a}. Assuming linear behavior of $\phi{}(T)$ in the temperature range from 5 to 15\,MK, we approximate $\phi{}(T) = AT+B$. Parameters $C$ and $A$ are estimated from the tabulated functions for T and EM calculation available in the SolarSoft package (the SXR flux per unit emission measure as a function of temperature). The results are averaged for the GOES10~--- GOES15 satellites (the difference is negligible). The derived parameter values are $C=0.7\times{}10^{-55}$\,erg\,cm\,s$^{-1}$, $A=2.22\times{}10^{-3}$\,K$^{-1}$. The flux equation becomes $F_{1-8{\AA}} = C\times{}EM\times{}(AT+B)$. At the EM maximum $\dfrac{dEM}{dt} = 0$, hence:
		
		\begin{gather}
		\label{eq:GOESDER}
		\dfrac{dF_{1-8{\AA}}}{dt} = A\times{}C\times{}EM\dfrac{dT}{dt}
		\end{gather}
		
		Replacing the temperature derivative from Eq.~\ref{eq:EBTEL-red1} we obtain:
		
		\begin{gather}
		\label{eq:GOESDER-eq2}
		\dfrac{dF_{1-8{\AA}}}{dt} = -\dfrac{1}{\sqrt{L}} \sqrt{EM\times{}S}\left(F_{0} + EM\dfrac{\Lambda{}}{2ST^{2/3}}\right) \dfrac{2AC(\gamma{}-1)}{k_{B}} \\
		\label{eq:looplength}
		L = EM\times{}S\left(\dfrac{dF_{1-8{\AA}}}{dt}\right)^{-2}\left(F_{0} + EM\dfrac{\Lambda{}}{2ST^{2/3}}\right)^{2} \dfrac{4A^{2}C^{2}(\gamma{}-1)^{2}}{k_{B}^{2}}
		\end{gather}
		
		There are two processes which can decrease the SXR flux: heat conduction at the loop footpoints and radiative losses along the loop. $F_{0} = \kappa{}_{0}T^{5/2}\dfrac{\partial{}T}{\partial{}s}|_{s=0}$, where $\kappa{}_{0}\approx{}10^{-6}$erg\,s$^{-1}$\,cm$^{-1}$\,K$^{-7/2}$ \citep{Spitzer53a}, represents the conduction flux at the loop footpoints. If the temperature increases from $T\approx{}10^{4}$\,K to $T\approx{}10^{6}$\,K at the loop footpoints on scales of $\approx{}1000$\,km, then $\dfrac{\partial{}T}{\partial{}s}\approx{}10^{-2}$\,K\,cm$^{-1}$, and flux in the transition region is $10^{7}$\,erg\,cm$^{-2}$s$^{-1}$. Notice that the radiative loss term is comparable with the conduction flux: for example, for the median characteristics of B6.3~- C1.6 class events $\dfrac{EM\times{}\Lambda}{2ST^{2/3}} \approx{}(2-3)\times{}10^{7}$\,erg\,cm$^{-2}$s$^{-1}$, where $\Lambda=10^{-17.73}$\,erg\,cm$^{3}$\,K\,s$^{-1}$, $T\approx{}(8-10)\times{}10^{6}$\,K, $EM=(8-9)\times{}10^{47}$\,cm$^{-3}$, $S\approx{}(0.7-0.8)\times{}10^{18}$\,cm$^{2}$ (half of the ribbon area).
		
\begin{figure}[h]
	\centering
	\includegraphics[width=0.49\linewidth]{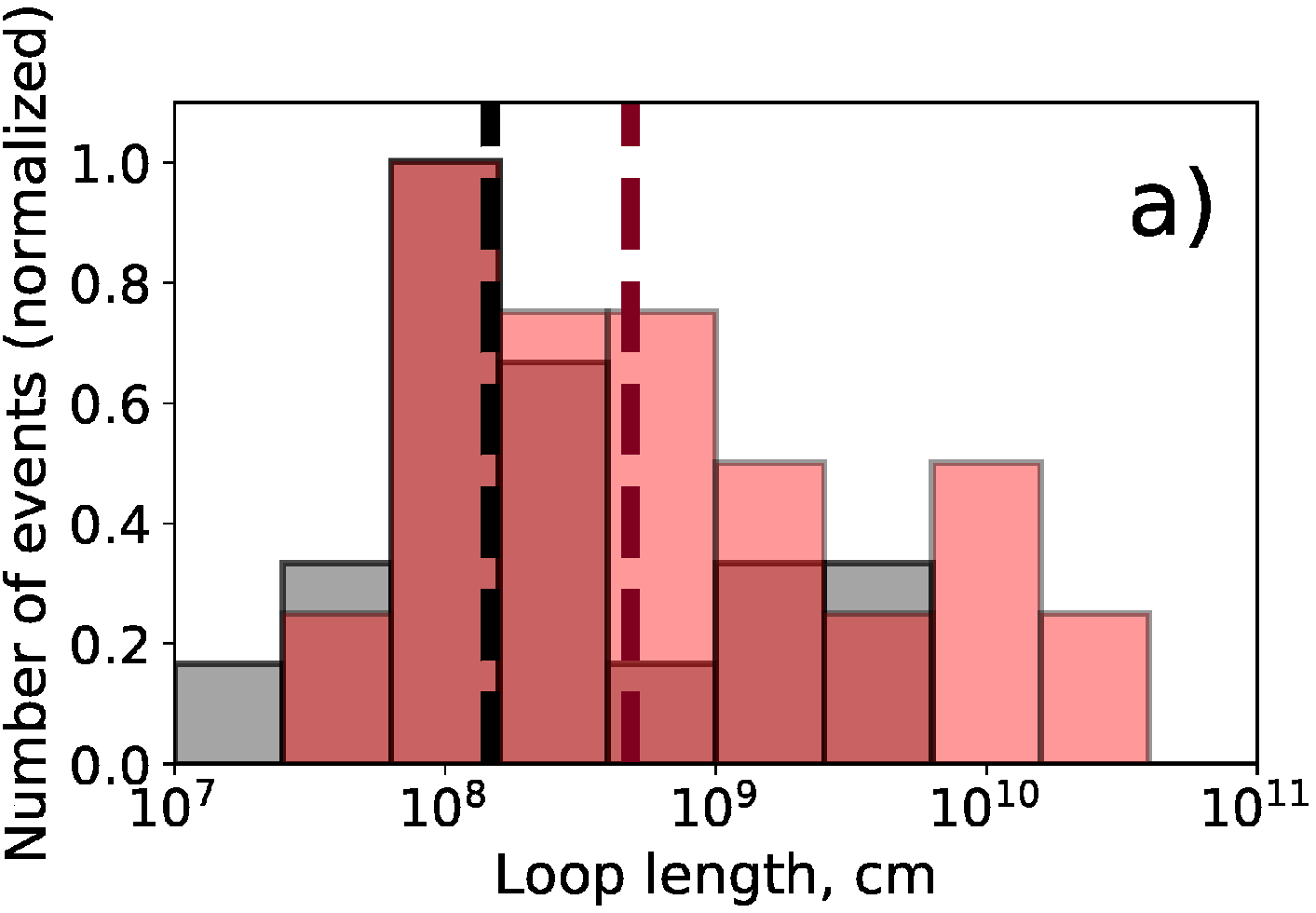}
	\includegraphics[width=0.49\linewidth]{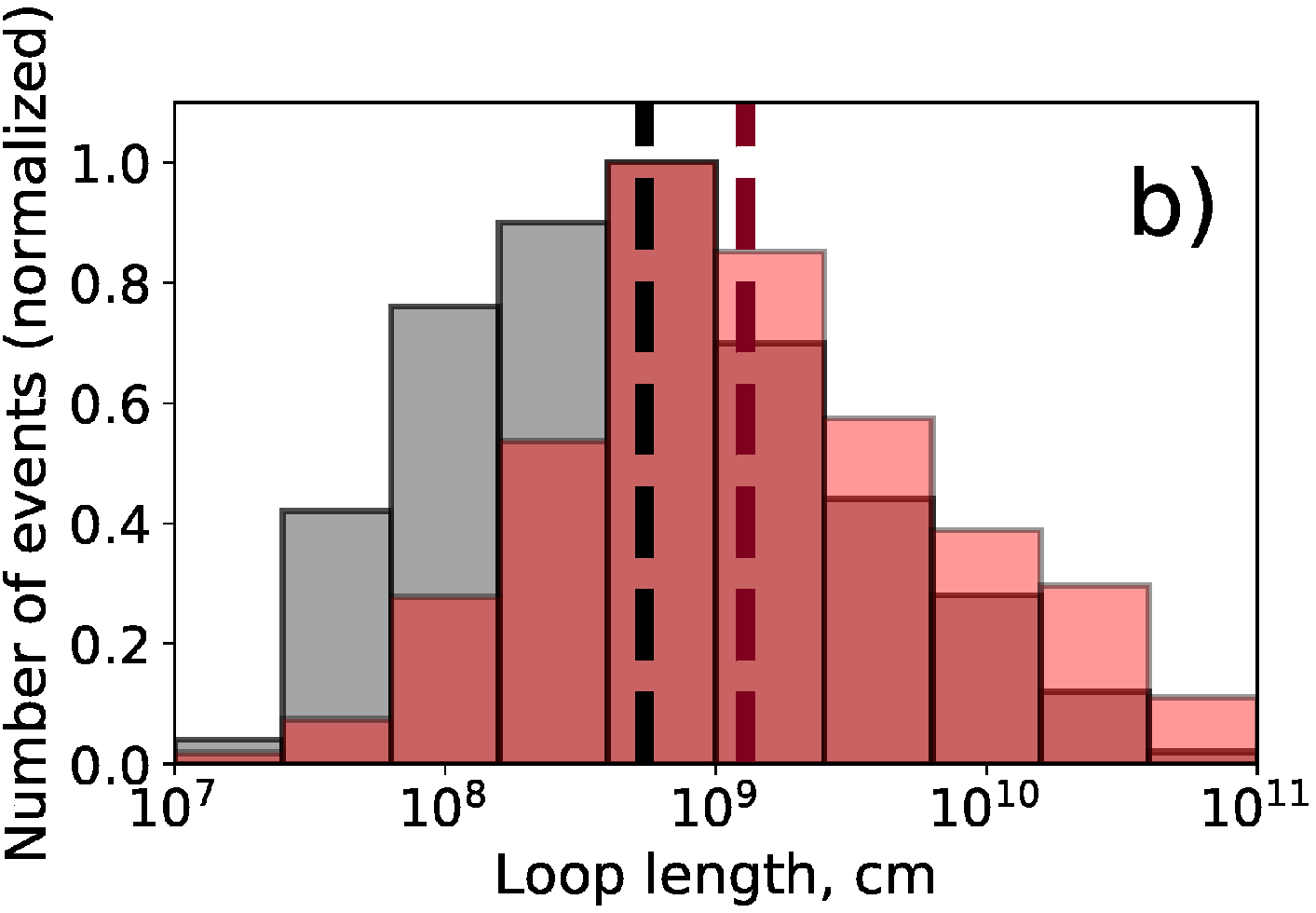} \\
	\includegraphics[width=0.49\linewidth]{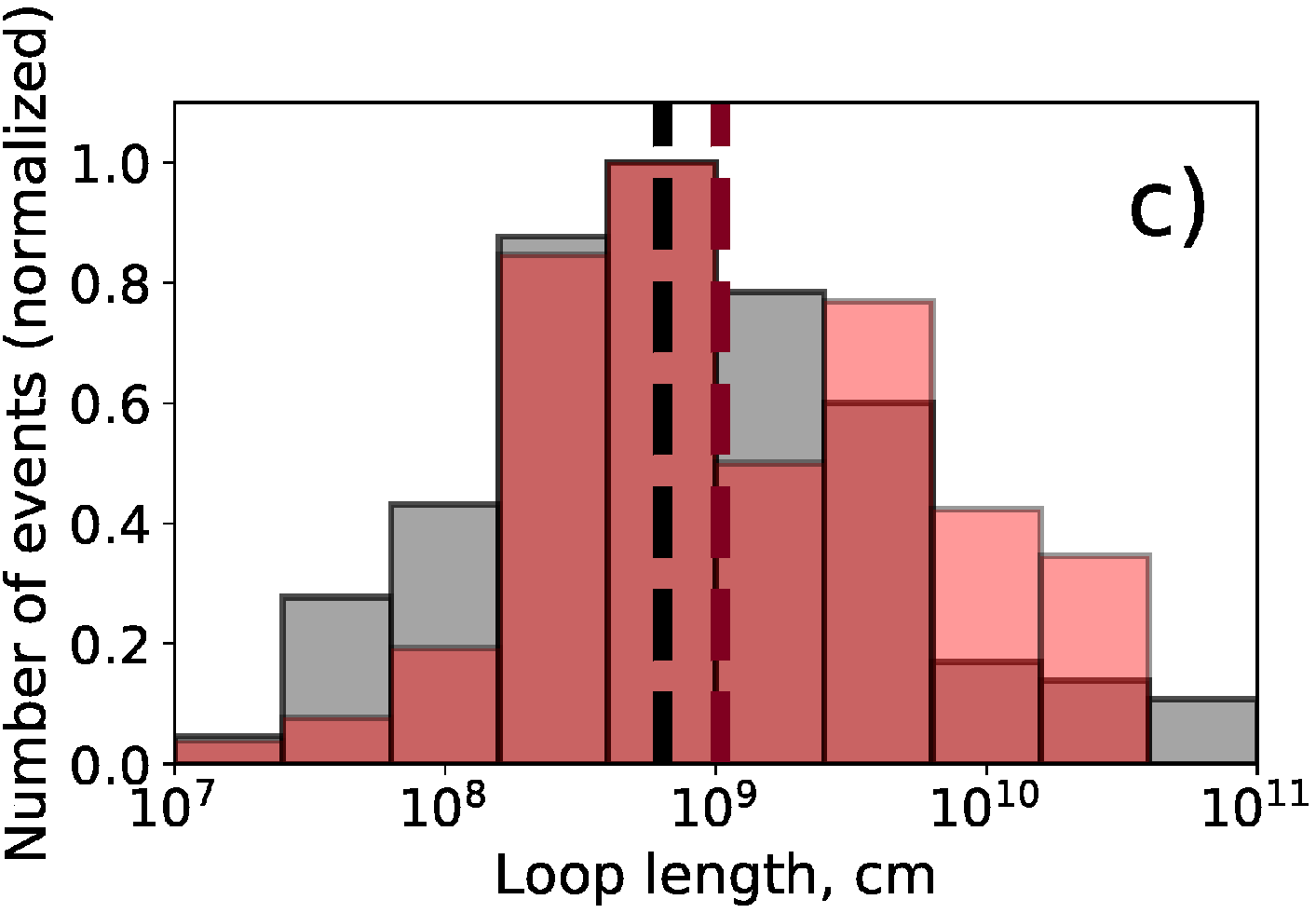}
	\includegraphics[width=0.49\linewidth]{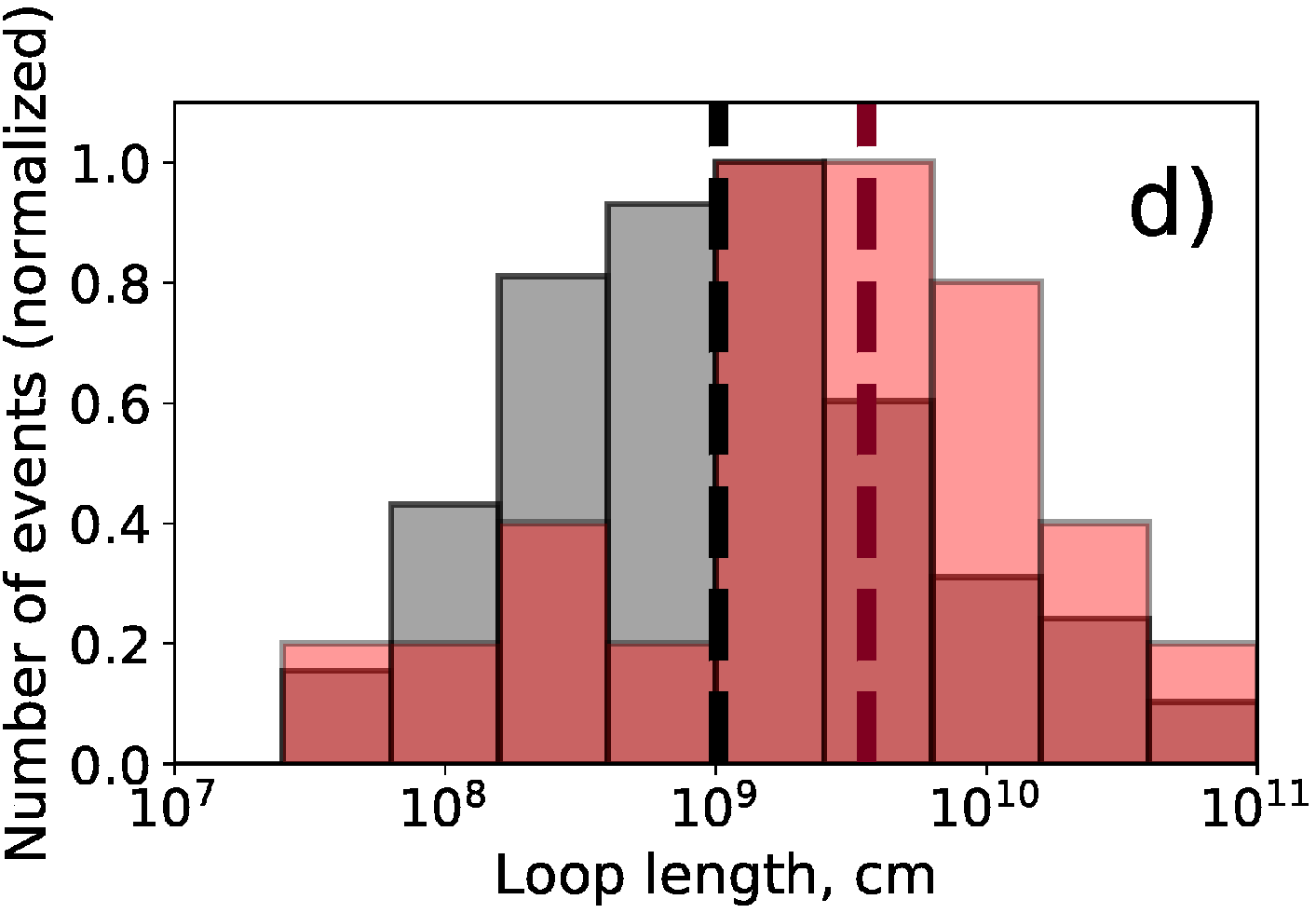} \\
	\caption{Histograms of the loop lengths calculated from Eq.~\ref{eq:looplength} (for $F_{0} = 10^{7}$\,erg\,cm$^{-2}$s$^{-1}$) for T-controlled (red) and EM-controlled (gray) events of (a) B2.5~- B6.3, (b) B6.3~- C1.6, (c) C1.6~- C4.0, and (d) C4.0~- M1.0 GOES class ranges. Red dashed vertical line represents the median value for T-controlled events, black~--- for EM-controlled events. The corresponding median absolute deviations are presented in Table~\ref{table2_looplengths}.}
	\label{figure:EBTELsupport}
\end{figure}

		Assuming $F_{0}=10^{7}$\,erg\,cm$^{-2}$s$^{-1}$ and the flaring loop cross-sections $S = S_{ribbon}/2$, we estimate from Eq.~\ref{eq:looplength} the lengths of the loops where the chromospheric evaporation was developed for every flare (for which information about the flare ribbon areas is available) in the B2.5~- B6.3, B6.3~- C1.6, C1.6~- C4.0, and C4.0~- M1.0 class ranges. The histograms for the selected class ranges are presented in Figure~\ref{figure:EBTELsupport}. The corresponding median values are presented in Table~\ref{table2_looplengths} (column for $F_{0}=10^{7}$\,erg\,cm$^{-2}$s$^{-1}$). The median values of the loop lengths are 2~- 4 times longer for the T-controlled flares in all selected class ranges than those for the EM-controlled events. Variation of the conduction flux from $10^{6}$\,erg\,cm$^{-2}$s$^{-1}$ to $10^{8}$\,erg\,cm$^{-2}$s$^{-1}$ leads to the same conclusion about the loop lengths (see Table~\ref{table2_looplengths}). However, high conduction fluxes ($\ge{}10^{8}$\,erg\,cm$^{-2}$s$^{-1}$) result in unrealistic median loop lengths of $\ge{}10^{10}$\,cm. The results allow us to conclude that the development of flares in coronal loops of different length can be one of the reasons for the difference between the T-controlled and EM-controlled events.
		
		In the case of two or more temporarily-overlapping distinct events of approximately the same magnitude occurring at the solar disk, the presented analysis is not valid. To estimate the influence of this effect, we analyzed spatially-resolved observations of flares obtained from the 131\,{\AA} data of the Atmospheric Imaging Assembly onboard Solar Dynamics Observatory \citep[SDO/AIA,][]{Lemen12a} using image processing algorithms \citep{Martens12}. These data are collected in the Heliophysics Event Knowledgebase \citep[HEK,][]{Hurlburt12}, which is incorporated in the IMIDSF \citep{Sadykov17a}. We found that in 2010-2017 about 6.5\% (530 out of 8115 events) of the GOES flares with the SDO/AIA 131\,{\AA} counterparts were overlapping with other flares of comparable EUV magnitude (at least a half of the EUV flux of the primary flare identified in the GOES catalog). However, the analysis of such overlapping events led us to qualitatively the same conclusions as for the entire data set (i.e. the existence of temporarily-overlapping flare events did not influence our conclusions).

\section{Summary and Conclusion}
\label{Section:conclusion}
	
	Our conclusions are the following:
	\begin{enumerate}
		\item The soft X-ray radiation of most flares (96.3\%) follows the sequential appearance of the temperature (T), radiation flux (SXR), and emission measure (EM) maxima in agreement with the chromospheric evaporation scenario. The fraction of such flares increases with the amplitude of the SXR maximum (GOES X-ray class). For 82.5\% of such flares, the SXR derivative reaches its maximum before the T maximum.
		\item The SXR maximum of weak B-class flares mainly occurs very close to the temperature maximum (34.0\% of the events are the T-controlled). The situation is opposite for the M- and X-class flares, 82.8\% of which are EM-controlled;
		\item The transition between the two regimes occurs in the range of X-ray class B6.3~- C1.6. The number of the T-controlled (1176) and EM-controlled (1365) events is almost the same in this class range;
		\item The following differences in the averaged physical parameters are found for the T-controlled and EM-controlled events (see Table~\ref{table1_TEBBSparameters}). Compared to the EM-controlled events, the T-controlled events have:
		\begin{itemize}
			\item[--] larger maximum emission measure and lower maximum temperature;
			\item[--] shorter relative growth time and longer duration;
			\item[--] larger SXR FWHM and longer characteristic decay times;
		\end{itemize}
		\item The lengths of the flare loops estimated from the analysis of a single loop dynamics supports the conclusion that the T-controlled events can be developed in longer loops than the EM-controlled events.
	\end{enumerate}
	
	This interpretation of the statistical results is based on several simplifications due to the limited information in the GOES data. It does not take into account the possible presence of heating in the gradual and decay phases of the flare \citep{Czaykowska99a,Czaykowska01a,Ryan13a} and the multi-thread nature of solar flares when the different coronal loops are heated at different times \citep{Warren06a}. Also, we consider a single-temperature approximation of the flaring plasma, although there is evidence for its multi-thermal structure \citep{Warmuth16a,Sharykin15a}. Further studies with the use of complimentary observations by RHESSI~\citep{Caspi14a}, SDO/AIA~\citep{Lemen12a,Ryan14a}, SDO/EVE~\citep{McTiernan18a}, SphinX/CORONAS-Photon~\citep{Gryciuk17a,Kirichenko17a}, etc. will increase understanding of the limitations of the single-temperature plasma approximation. Nevertheless, the statistical analysis of 14955 events performed in this study allows us to better understand the physical characteristics of flare SXR emission.

\acknowledgments

We acknowledge the GOES and SDO/AIA teams for the availability of the high-quality scientific data. The research was partially supported by the NASA Grants NNX12AD05A, NNX14AB68G and NNX16AP05H, and NSF grant 1639683.

\bibliographystyle{aasjournal}

\bibliography{TEBBSstatistics}
\clearpage

\clearpage

\clearpage

\clearpage

\clearpage

\clearpage

\clearpage

\begin{sidewaystable}
	\centering
	\caption{Median values and median absolute deviations of the SXR characteristics of the T-controlled and EM-controlled flares for different GOES class ranges. The last column gives the number of events for which information about the flare ribbon areas is available \citep{Kazachenko17a}.}
	\label{table1_TEBBSparameters}
	\tiny
	\begin{tabular}{|c|c|c|c|c|c|c|c|c|c|}
		\hline
		\multirow{2}{*}{GOES class}	&	\multirow{2}{*}{Regime}	&	Number of	&	\multicolumn{2}{c|}{Physical characteristics}	&	\multicolumn{4}{c|}{Flare temporal characteristics, min}	&	Number of events in	\\
		\cline{4-9}
			&		&	events	&	T max, $10^{6}$\,K	&	EM max, $10^{48}\,$cm$^{-3}$	&	Duration, min	&	Relative growth time, min	&	FWHM, min	&	Decay time, min	&	Flare Ribbon Catalog	\\
		\hline
		\multirow{2}{*}{B1.0~- B2.5}	&	T-controlled	&	329		&	9.87$\pm$0.71	&	0.23$\pm$0.06	&	13.2$\pm$5.0	&	0.43$\pm$0.09	&	9.20$\pm$4.00	&	3.79$\pm$1.12	&	0	\\
										&	EM-controlled	&	198		&	10.9$\pm$1.1	&	0.17$\pm$0.04	&	11.2$\pm$4.9	&	0.57$\pm$0.10	&	6.98$\pm$3.43	&	3.33$\pm$1.39	&	0	\\
		\hline
		\multirow{2}{*}{B2.5~- B6.3}	&	T-controlled	&	946		&	10.6$\pm$0.9	&	0.43$\pm$0.09	&	15.2$\pm$6.3	&	0.44$\pm$0.09	&	10.6$\pm$4.2	&	6.44$\pm$2.38	&	19	\\
										&	EM-controlled	&	639		&	11.1$\pm$0.9	&	0.36$\pm$0.07	&	11.8$\pm$5.1	&	0.60$\pm$0.09	&	7.65$\pm$3.65	&	5.06$\pm$2.43	&	18	\\
		\hline
		\multirow{2}{*}{B6.3~- C1.6}	&	T-controlled	&	1176	&	11.2$\pm$0.9	&	0.89$\pm$0.22	&	20.4$\pm$10.2	&	0.42$\pm$0.09	&	14.2$\pm$6.6	&	9.20$\pm$4.64	&	228	\\
										&	EM-controlled	&	1365	&	12.0$\pm$1.1	&	0.80$\pm$0.18	&	13.5$\pm$5.9	&	0.58$\pm$0.09	&	8.65$\pm$4.30	&	6.00$\pm$3.33	&	240	\\
		\hline
		\multirow{2}{*}{C1.6~- C4.0}	&	T-controlled	&	660		&	12.0$\pm$1.1	&	1.80$\pm$0.38	&	23.1$\pm$10.9	&	0.42$\pm$0.10	&	16.3$\pm$7.6	&	6.49$\pm$2.98	&	112	\\
										&	EM-controlled	&	1434	&	13.3$\pm$1.1	&	1.69$\pm$0.34	&	15.0$\pm$6.4	&	0.58$\pm$0.09	&	9.73$\pm$4.88	&	5.83$\pm$2.87	&	289	\\
		\hline
		\multirow{2}{*}{C4.0~- M1.0}	&	T-controlled	&	155		&	13.3$\pm$1.2	&	3.77$\pm$0.73	&	32.6$\pm$14.7	&	0.40$\pm$0.13	&	24.2$\pm$11.8	&	11.2$\pm$4.35	&	23	\\
										&	EM-controlled	&	1138	&	14.9$\pm$1.2	&	3.67$\pm$0.74	&	16.9$\pm$7.4	&	0.59$\pm$0.10	&	10.6$\pm$5.4	&	8.12$\pm$4.19	&	273	\\
		\hline
		\multirow{2}{*}{$\ge$M1.0}		&	T-controlled	&	21		&	16.0$\pm$1.1	&	8.37$\pm$1.38	&	31.5$\pm$16.0	&	0.50$\pm$0.18	&	23.6$\pm$15.3	&	8.90$\pm$5.23	&	4	\\
										&	EM-controlled	&	1030	&	17.5$\pm$1.6	&	11.1$\pm$4.5	&	20.5$\pm$9.6	&	0.61$\pm$0.09	&	11.4$\pm$6.0	&	8.27$\pm$4.39	&	215	\\
		\hline
	\end{tabular}
\end{sidewaystable}
\clearpage

\begin{sidewaystable}
	\centering
	\caption{Median values and median absolute deviations of the SXR flux derivative and temperature during EM maximum, flare ribbon area, and loop length of the T-controlled and EM-controlled flares for different GOES class ranges. Loop lengths are is estimated from Eq.~\ref{eq:looplength} for every event for each of three values of conduction flux, $F_{0}$, presented in the table.}
	\label{table2_looplengths}
	\tiny
	\begin{tabular}{|c|c|c|c|c|c|c|c|}
		\hline
		\multirow{2}{*}{GOES class}	&	\multirow{2}{*}{Regime}	&	Flux derivative during	&	Temperature during	&	Flare ribbon	&	\multicolumn{3}{c|}{Loop lengths, $10^{8}$\,cm}	\\
		\cline{6-8}
			&		&	EM maximum, 10$^{-9}$\,W\,m$^{-2}$s$^{-1}$	&	EM maximum, MK	&	area, 10$^{18}$\,cm$^{2}$	&	$F_{0} = 10^{6}$\,erg\,cm$^{-2}$s$^{-1}$	&	$F_{0} = 10^{7}$\,erg\,cm$^{-2}$s$^{-1}$	&	$F_{0} = 10^{8}$\,erg\,cm$^{-2}$s$^{-1}$	\\
		\hline
		\multirow{2}{*}{B2.5~- B6.3}	&	T-controlled	&	0.55$\pm$0.30	&	7.11$\pm$0.95	&	1.04$\pm$0.50	&	1.32$\pm$1.14	&	4.82$\pm$3.82	&	104.4$\pm$78.7	\\
										&	EM-controlled	&	0.74$\pm$0.41	&	9.57$\pm$0.97	&	0.99$\pm$0.22	&	0.63$\pm$0.44	&	1.46$\pm$1.07	&	41.2$\pm$31.5	\\
		\hline
		\multirow{2}{*}{B6.3~- C1.6}	&	T-controlled	&	0.97$\pm$0.58	&	7.79$\pm$1.02	&	1.36$\pm$0.42	&	6.18$\pm$5.42	&	12.9$\pm$10.7	&	189.3$\pm$154.2	\\
										&	EM-controlled	&	1.40$\pm$0.81	&	10.48$\pm$1.13	&	1.60$\pm$0.50	&	2.34$\pm$2.03	&	5.44$\pm$4.50	&	85.4$\pm$71.5	\\
		\hline
		\multirow{2}{*}{C1.6~- C4.0}	&	T-controlled	&	2.06$\pm$1.20	&	8.76$\pm$1.31	&	2.45$\pm$0.78	&	6.20$\pm$5.21	&	10.4$\pm$8.5	&	146.0$\pm$116.3	\\
										&	EM-controlled	&	2.70$\pm$1.36	&	11.59$\pm$1.21	&	2.27$\pm$0.69	&	3.31$\pm$2.70	&	6.33$\pm$5.21	&	86.0$\pm$69.9	\\
		\hline
		\multirow{2}{*}{C4.0~- M1.0}	&	T-controlled	&	3.26$\pm$1.91	&	9.89$\pm$1.48	&	4.01$\pm$1.58	&	24.8$\pm$18.4	&	36.0$\pm$27.9	&	411.4$\pm$359.6	\\
										&	EM-controlled	&	5.01$\pm$2.48	&	13.08$\pm$1.03	&	3.57$\pm$1.02	&	6.70$\pm$5.53	&	10.2$\pm$8.3	&	98.3$\pm$82.1	\\
		\hline
	\end{tabular}
\end{sidewaystable}

\end{document}